\documentclass[journal=nalefd,manuscript=letter,layout=traditional,super=true]{achemso}
\usepackage[version=3]{mhchem} % Formula subscripts using \ce{}
\usepackage{bm}% bold math
\usepackage{float}
\usepackage{braket}
\usepackage[colorlinks=true, allcolors=blue, linktocpage=true]{hyperref}

\title{All-optical Tuning of Indistinguishable Single-Photons Generated in Three-level Quantum Systems}
	
\author{\L{}ukasz Dusanowski}
\email{lukaszd@princeton.edu}
\affiliation{Technische Physik and W\"{u}rzburg-Dresden Cluster of Excellence ct.qmat, Physikalisches Institut and Wilhelm-Conrad-R\"{o}ntgen-Research Center for Complex Material Systems, University of W\"{u}rzburg, Am Hubland, D-97074 W\"{u}rzburg, Germany}	
\altaffiliation{Present address: Department of Electrical and Computer Engineering, Princeton University, Princeton, NJ 08544, USA}

\author{Chris Gustin}
\affiliation{Department of Applied Physics, Stanford University, Stanford, California 94305, USA}
\altaffiliation{Department of Physics, Engineering Physics, and Astronomy, Queen's University, Kingston, Ontario K7L 3N6, Canada}

\author{Stephen Hughes}
\affiliation{Department of Physics, Engineering Physics, and Astronomy, Queen's University, Kingston, Ontario K7L 3N6, Canada}

\author{Christian Schneider}
\affiliation{Technische Physik and W\"{u}rzburg-Dresden Cluster of Excellence ct.qmat, Physikalisches Institut and Wilhelm-Conrad-R\"{o}ntgen-Research Center for Complex Material Systems, University of W\"{u}rzburg, Am Hubland, D-97074 W\"{u}rzburg, Germany}
\altaffiliation{Institute of Physics, University of Oldenburg, D-26129 Oldenburg, Germany}

\author{Sven H\"{o}fling}
\affiliation{Technische Physik and W\"{u}rzburg-Dresden Cluster of Excellence ct.qmat, Physikalisches Institut and Wilhelm-Conrad-R\"{o}ntgen-Research Center for Complex Material Systems, University of W\"{u}rzburg, Am Hubland, D-97074 W\"{u}rzburg, Germany}

\date{\today}

\begin{document}
	
	%\pacs{85.35.Be, 42.50.Ar, 42.82.-m, 42.50.Dv}
	
	%\keywords{single photon source, Mollow triplet, Autler-Townes effect, resonance fluorescence, quantum dot, two-photon interference}
	
	\begin{abstract} 
		Resonance fluorescence of two-level quantum systems has emerged as a powerful tool in quantum information processing. Extension of this approach to higher-level systems provides new opportunities for quantum optics applications. Here we introduce a coherent driving scheme of a three-level ladder system utilizing Autler-Townes and ac Stark effects by resonant excitation with two laser fields. We propose theoretically and demonstrate experimentally the feasibility of this approach towards all-optical spectral tuning of quantum dot-based single-photon sources and investigate photon indistinguishability and purity levels. Our tuning technique allows for fast optical control of the quantum emitter spectrum which paves the way towards temporal and spectral shaping of the single photons, formation of topological Floquet states or generation of high-dimensional frequency-encoded quantum states of light.
	\end{abstract}
	
	\maketitle

	The quantum mechanical interaction of light and matter underpins a broad spectrum of physical phenomena and their applications in quantum information processing. One example is Rabi oscillations, where the occupation probability of a two-level quantum system cycles between 0 and 1 when coherently coupled to a resonant field. By driving its dipole moment with coherent optical pulses, the two-level quantum state can be arbitrarily rotated on the Bloch sphere, which is the hallmark of coherent qubit control, utilized in various systems such as atoms~\cite{Gibbs1972}, quantum dots (QDs)~\cite{Zrenner2002,Xiaoqin2003}, defects in solids~\cite{Togan2010} and superconducting circuits~\cite{Astafiev2010}. When a two-level system is driven by a continuous wave (cw) strong resonant drive, an emission spectrum composed of three resonance peaks, known as the Mollow triplet~\cite{Mollow1969,Schuda1974,Baur2009,Flagg2009}, is observed. This effect can be explained by the optical transitions within a dressed-state picture, where the system eigenstates hybridize light-matter components, leading to an energy splitting equal to the Rabi energy.

	Generalization of these phenomena to three-level quantum systems driven by two non-degenerate light fields leads to a variety of other physical effects including  Autler-Townes (AT) splitting~\cite{Autler1955,Xu2007,Zhou2017}, electromagnetically induced transparency~\cite{Boller1991,Chaneliere2005,Brunner2009}, control of exciton fine-structure in QDs~\cite{Muller2009}, dressing the biexciton~\cite{Muller2008}, coherent control of spin~\cite{Press2008} and non-linear absorption~\cite{Gerardot2009,Nguyen2018}. Furthermore, resonant spectroscopy has proven to be a powerful technique for the generation of single photons with high purity and mutual indistinguishability~\cite{Maunz2007,He2013a}.
	
	Utilizing these effects for optical control of the quantum emitter frequency is particularly appealing as a step towards fast frequency modulation of two-level systems~\cite{Silveri2017}, which could enable single photon pulse shaping\cite{Heinze2015}, creation of high-dimensional frequency-encoded quantum states of light~\cite{Shlesinger2021}, generation of exotic single-photon spectrum or formation of topological Floquet states~\cite{Lukin2020}.  
	
	In this work, we explore both experimentally and theoretically the possibility of using a three-level ladder system driven by continuous and pulsed non-degenerate resonant fields, for frequency- and time-domain control of single-photon generation. Specifically, using the biexciton-exciton radiative cascade in a semiconductor QD, we show the feasibility of this driving scheme for the generation of frequency tunable single photons and investigate efficiency, photon purity and indistinguishability levels achievable using this approach.
	
\begin{figure}[t]
		\centering
		\includegraphics[width=0.6\columnwidth]{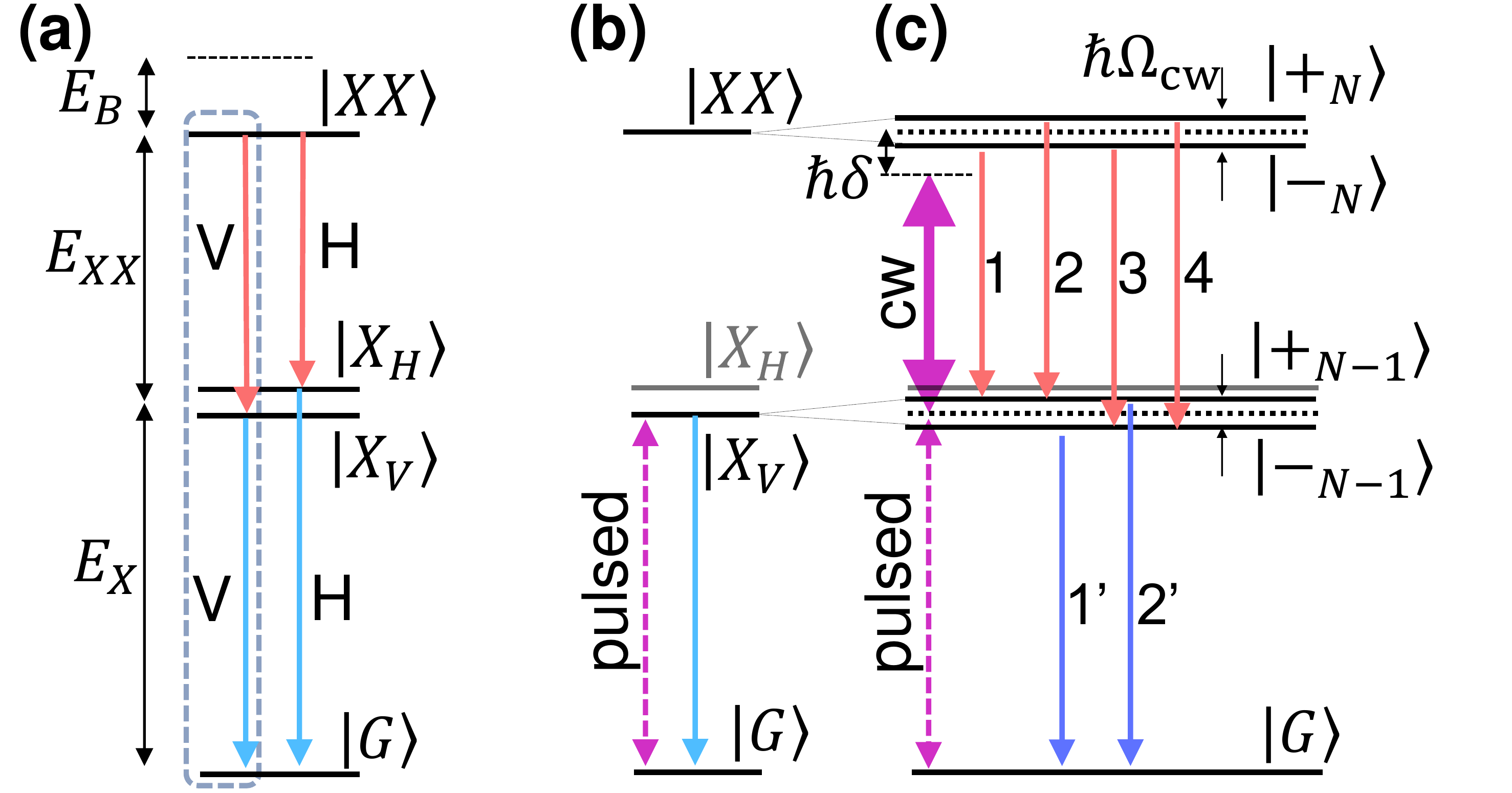}
		\caption{\label{fig:1}
			\textbf{Dressed three-level ladder system scheme.} \textbf{(a)},~Bare biexciton-exciton cascade. \textbf{(b)},~The excitonic $\ket{X_V}$-$\ket{G}$ transition is driven resonantly by a V-polarized pulsed excitation laser, followed by subsequent emission of a single photon. \textbf{(c)},~A second V-polarized laser, operating in cw mode tuned to the $\ket{XX}$-$\ket{X_V}$ transition, perturbs the system and dresses the bare $\ket{XX}$ and $\ket{X_V}$ levels into pairs of hybrid two-level/field states. The resulting emission spectra consists of 4-transitions originating from the biexciton (1-4) forming a Mollow triplet, and two transitions from the exciton (1',2') forming an AT doublet.	
		}
	\end{figure}

	In Fig.~\ref{fig:1}(a), we show an energy level diagram of the biexciton-exciton cascade for a QD. It consists of four states: biexciton $|XX\rangle$ (two electron-hole pairs), two fine-structure split excitonic states $|X_V\rangle$ and $|X_H\rangle$ (single electron-hole pairs with different spin configurations), and the QD ground state $|G\rangle$. Radiative recombination of one electron-hole pair from $|XX\rangle$ results in the emission of a single vertical (V) or horizontal (H) linearly polarized photon, and a change of the QD state into $|X_V\rangle$ or $|X_H\rangle$, correspondingly. Similarly, in the case of $|X_{V/H}\rangle$-$|G\rangle$ radiative transitions, single photons with V/H polarization are emitted. Note that emitted V (H) polarized photons from $|XX\rangle$ and $|X_{V/H}\rangle$ are polarization-entangled~\cite{Akopian2006,Salter2010,Jayakumar2014,Prilmuller2018}.
	
    Here we focus on the V-polarized cascade, which can be implemented by polarization-selective QD excitation and emission filtering. We achieve this by using an integrated waveguide (WG) device with an embedded QD~\cite{Dusanowski2019}, aligned along the [110] GaAs crystallographic axis, which conveniently corresponds to the H-polarization orientation of the QD dipole. Thus, part of the V-polarized emission couples into the transverse-electric mode of the WG, where it is guided around 1~mm on-chip until being collected by the sample side facet. Light out-coupled from the WG exhibits a linear polarization degree of over 99\%. More details can be found in Supporting Information (SI) Sections 2 and 3. Excitation of the QD is performed from the top of the WG, allowing pumping at any desired wavelength and in-plane polarization. Importantly, this pumping scheme allows for efficient suppression of the laser scattered on the sample. 
	
	For our experiments on the QD cascade, we introduce two driving fields: (i)~a V-polarized pulse resonant with the  $|X_{V}\rangle$-$|G\rangle$ transition, and (ii)~a V-polarized cw drive (near)-resonant with $|XX\rangle$-$|X_V\rangle$, as shown in Fig.~\ref{fig:1}(c). For a strong cw drive, the $|XX\rangle$-$|X_V\rangle$ transition is dressed by the laser, resulting in an energy level splitting into hybrid $\ket{XX}$-$\ket{X_V}$-light states, denoted as $|+_N\rangle$, $|-_N\rangle$ and $|+_{N-1}\rangle$, $|-_{N-1}\rangle$, where $N$ is the laser field photon number (using a Fock state basis). The emission spectra consists of six transitions, four of which are related to the dressed biexciton transition (1-4) forming a Mollow triplet, and two related to the exciton-ground transition (1',2') forming an AT doublet.

    In a rotating frame and without considering the pulse drive, the relevant system Hamiltonian is
	$H_S {=} \hbar\delta\sigma^{XX}_z/2 {+} {\hbar\Omega_{\rm cw}}\sigma_x^{XX}/2$, where $\Omega_{\rm cw}$ is the cw drive Rabi frequency, $\delta$ is the detuning of cw drive from $|XX\rangle$-$|X_V\rangle$ transition such that $\omega_{\rm cw} {=} E_{XX}/\hbar {-} \delta$, $\sigma_x^{XX} {=} \ket{X_V}\bra{XX} {+} \ket{XX}\bra{X_V}$, and $\sigma_z^{XX} {=} \ket{XX}\bra{XX} {-} \ket{X_V}\bra{X_V} $ (see SI Section 4; Fig.~\ref{fig:1}). The Hamiltonian eigenenergies are $E_{\pm} {=} \pm\hbar\eta/2$, where $\eta {=} \sqrt{\Omega_{\rm cw}^2 {+} \delta^2}$, with dressed states
	\begin{equation}
	\ket{\pm} = \frac{1}{\sqrt{\Omega_{\rm cw}^2+(\delta \mp \eta)^2}}\left[\Omega_{\rm cw}\ket{XX} - (\delta \mp \eta)\ket{X_V}\right].
	\end{equation}  
	As shown in Fig.~\ref{fig:1}(c), this system gives rise to optical transitions at energies $E_{XX} {-} \hbar\delta$, $E_{XX} {-}\hbar(\delta {\pm} \eta)$ (biexciton Mollow triplet), as well as $E_X {+} \hbar(\delta
	{\pm} \eta)/2$ (exciton AT doublet). For $|\delta| \gg \Omega_{\rm cw}$, the exciton emission approaches the ac Stark shift regime, and one of its doublet transitions dominates at  $E_{X} {-}  \hbar\Omega_{\rm cw}^2/(4\delta)$. We define the ac Stark shift $\hbar\Delta_{\rm ac}$ by subtracting the shifted exciton resonance from its unperturbed (bare) energy.  Thus, by changing the cw laser power one controls the AT (or ac Stark) frequency splitting, and by changing the power of the pulsed laser field, one can coherently control the ground versus  excited dressed state population.
	
    For our optical experiments, we identified a single QD embedded in the WG structure with a biexciton binding energy $E_B$ of 3.24~meV. Polarization-resolved experiments reveal an exciton fine-structure splitting of $13.6{\pm}0.2~\mu$eV and polarization alignment with the WG orientation (see SI Section 3). We define V-polarization as orthogonal to the WG. Next, for this particular QD, we dress the $|XX\rangle$-$|X_V\rangle$ transition; we perform photoluminescence (PL) measurements under above-band (incoherent) cw excitation, allowing observation of exciton and biexciton complexes simultaneously, and then introduce a second cw laser beam resonant with $|XX\rangle$-$|X_V\rangle$ and polarized along the V-direction. To model this system (including the pulsed drive) and calculate emission spectra and other observables, we use a quantum master equation. The coupling of excitons to longitudinal acoustic phonons can have significant effects on optically-driven semiconductor QDs, and so for numerical calculations we use a phonon master equation model adapted from Refs.~\cite{Gustin2019,hargart16} (more details in SI Section 4).  	

    \begin{figure}
		\centering
		\includegraphics[width=0.6\columnwidth]{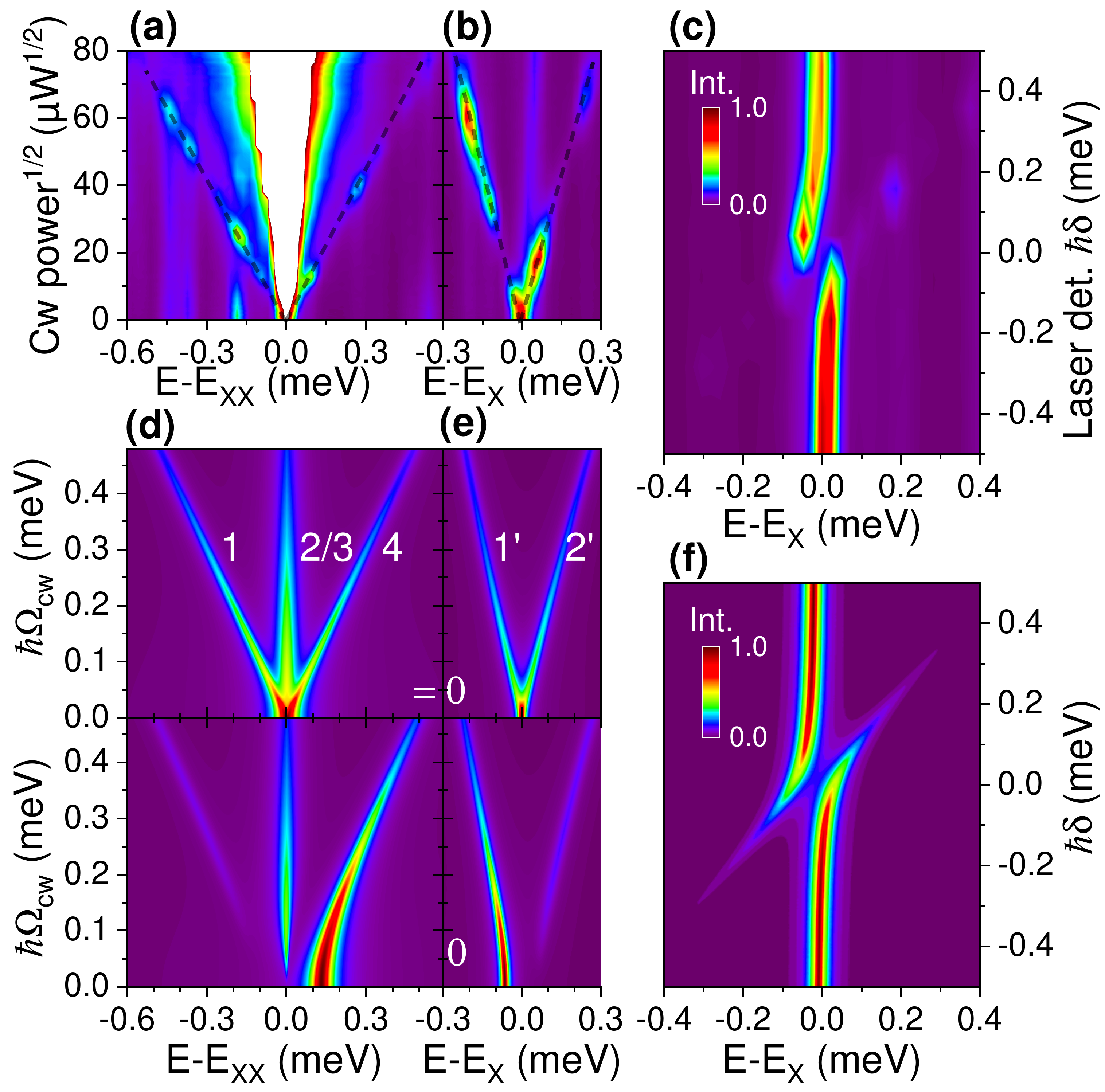}
		\caption{\label{fig:2}\textbf{Exciton and biexciton emission with variable dressing field intensity.} \textbf{(a)},\textbf{(b)}, Color-scale emission intensity map for exciton and biexciton emission vs square root of cw driving laser power incident on $\ket{XX}$-$\ket{X_V}$ transition. White color region in (a) is related to the leaking excitation laser. \textbf{(c)},~Emission intensity map for exciton vs cw driving laser detuning from $\ket{XX}$-$\ket{X_V}$ transition. \textbf{(d)},\textbf{(e)},\textbf{(f)},~Corresponding theoretically calculated emission spectra for on resonance $\delta=0$ (top d,e panels) and detuning $\hbar\delta= 130 \ \mu$eV (bottom panels).
		}
	\end{figure}
	
	\begin{figure}
		\centering
		\includegraphics[width=1\columnwidth]{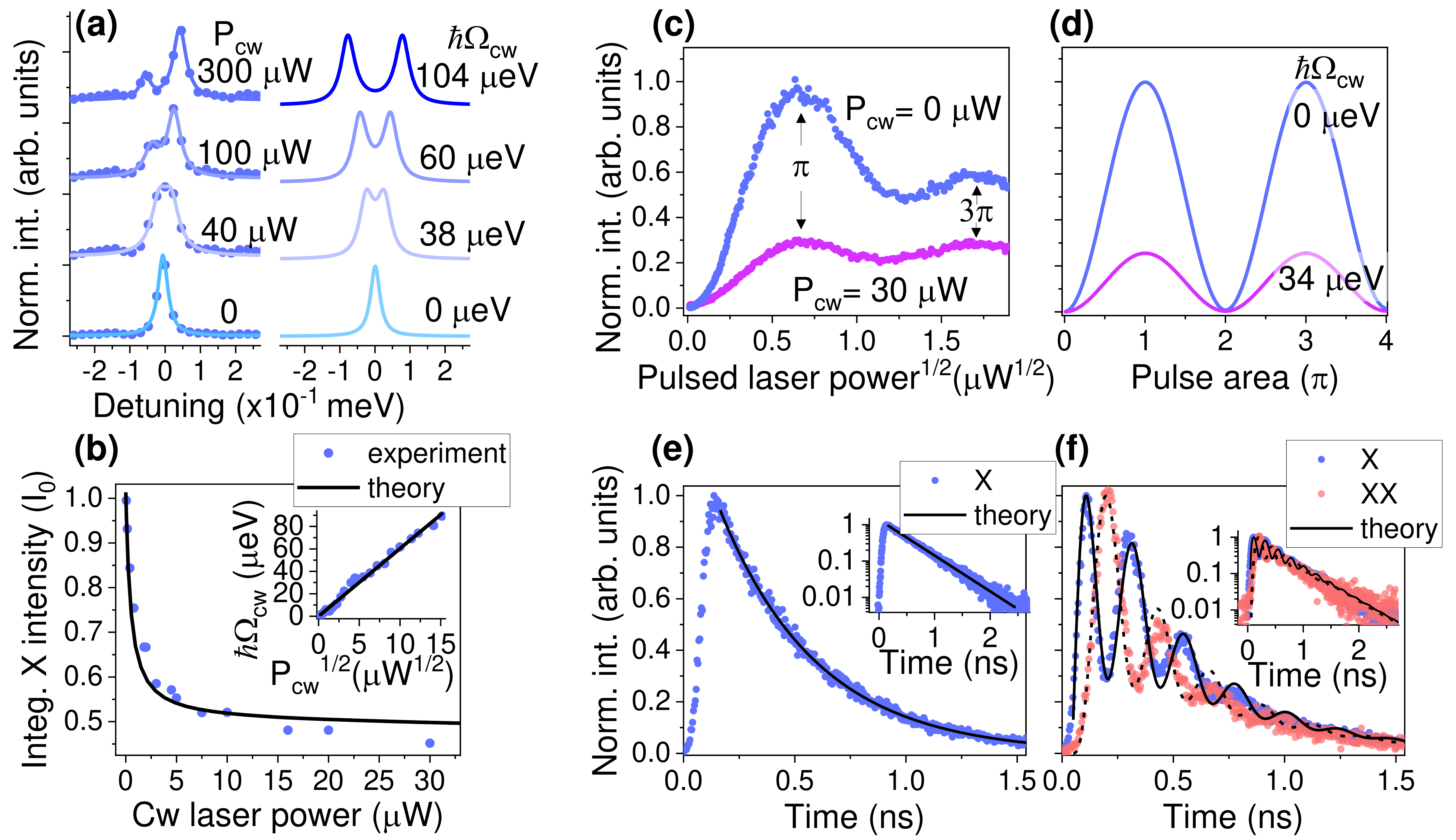}
		\caption{\label{fig:3}\textbf{Autler-Townes doublet.} \textbf{(a)},~Normalized resonance fluorescence spectra of the exciton for increasing  $\ket{XX}$-$\ket{X_V}$ incident cw pump power ($\delta=0$), showing the effect of the dressing field on the emission spectrum. Corresponding theoretically calculated emission spectra is shown in the right. \textbf{(b)},~Integrated exciton intensity as a function of the cw laser power. Inset: AT exciton doublet splitting vs square root of cw laser power (black solid lines represent theoretical expectation). \textbf{(c)},~Rabi oscillations of the exciton emission intensity as a function of the square root of the pulsed laser power for bare (blue points) and dressed (magenta points) case. \textbf{(d)},~Theoretically calculated Rabi oscillations of exciton for corresponding cw dressing strengths $\hbar\Omega_{\rm cw}$.  \textbf{(e)},~Normalized resonance fluorescence time-trace of the bare exciton transition under $\pi$-pulse excitation, exhibiting mono-exponential decay with 500~ps time constant ($\hbar\gamma_X = 1.32 \ \mu$eV). \textbf{(f)},~Time-trace of the exciton and biexciton transitions dressed with 30~$\mu$W power cw pump. Exciton and biexciton decays are fitted with Eq.~\ref{eq:pop} convolved with the setup IRF (100~ps width), resulting in a Rabi energy of $\hbar\Omega_{\rm cw}$~=~18~$\mu$eV. Insets in c,d: Semi-log plots of the corresponding graphs. 
		}
	\end{figure}

    In Figs.~\ref{fig:2}(a),(b), and (d),(e) we show, respectively, experimental and theoretical color-coded PL spectra as a  function of the square root of the dressing cw laser power (Rabi energy) for the biexciton and exciton emission. For the resonant drive $\delta$=0 case, as the dressing laser intensity increases, the biexciton (exciton) emission line splits into a Mollow triplet (AT doublet). The experimental spectra are modulated in frequency, caused by reflections from the WG facets forming a Fabry-P\'erot cavity~\cite{Dusanowski2019} (free-spectral-range of 0.2~meV and broadening of 30~$\mu$eV). In Fig.~\ref{fig:2}(c), we show the exciton emission spectra for a fixed pumping power of 250~$\mu$W as a function of laser detuning from the $|XX\rangle$-$|X_V\rangle$ transition. Corresponding theoretical spectra is shown in Fig.~\ref{fig:2}(f). The dressed AT exciton emission exhibits a characteristic avoided crossing, while for large detunings transitions asymptotically approach the non-perturbed energies. The remaining energy offset is the ac Stark shift. In the lower panels of Figs.~\ref{fig:2}(d),(e) we plot theoretically calculated spectra for the fixed detuning of $\hbar\delta$~=~130~$\mu$eV demonstrating asymmetry in the split doublets with dominating transitions 4 and 1'.
	
    Next, we switch to a fully coherent excitation scheme, corresponding to Fig.~\ref{fig:1}(c).  We resonantly drive the $|X_V\rangle$-$|G\rangle$ transition with a 2~ps full width at half maximum laser pulse, and the $|XX\rangle$-$|X_V\rangle$ transition with a cw laser, both polarized along the V-direction. The resulting normalized exciton resonance fluorescence spectra as a function of the cw laser power are shown in Fig.~\ref{fig:3}(a), and the corresponding theoretically calculated spectra are plotted in b. The exciton emission exhibits clear AT splitting at higher powers. An intensity imbalance within the doublet is observed in the experiments compared to theory, which we attribute to the Fabry-P\'erot cavity variation of the optical density of states (which is assumed to be constant in the theoretical model). With increasing cw laser power, the intensity of the doublet drops down until it reaches approximately half of its initial value [see Fig.~\ref{fig:3}(b)], as the strongly dressed exciton-biexciton system has nearly equal decay rates through both polarization channels. 

    In Fig.~\ref{fig:3}(c), we plot the exciton emission intensity versus the square root of pulsed laser power for bare exciton (blue points) and single (higher energy) AT split line at 30~$\mu$W cw laser power. Clear oscillatory behavior is observed in both cases related to coherent control of the exciton population. Oscillations are slightly damped partially caused by phonon-induced damping~\cite{Forstner2003,Gustin2018}. In the dressed case, an approximate four-fold decrease of split line intensity under $\pi$-pulse excitation is observed, due to AT splitting and the aforementioned population transfer. Figure~\ref{fig:3}(d) shows the corresponding simulated exciton emitted photon number.

	\begin{figure}[htb]
		\centering
		\includegraphics[width=0.6\columnwidth]{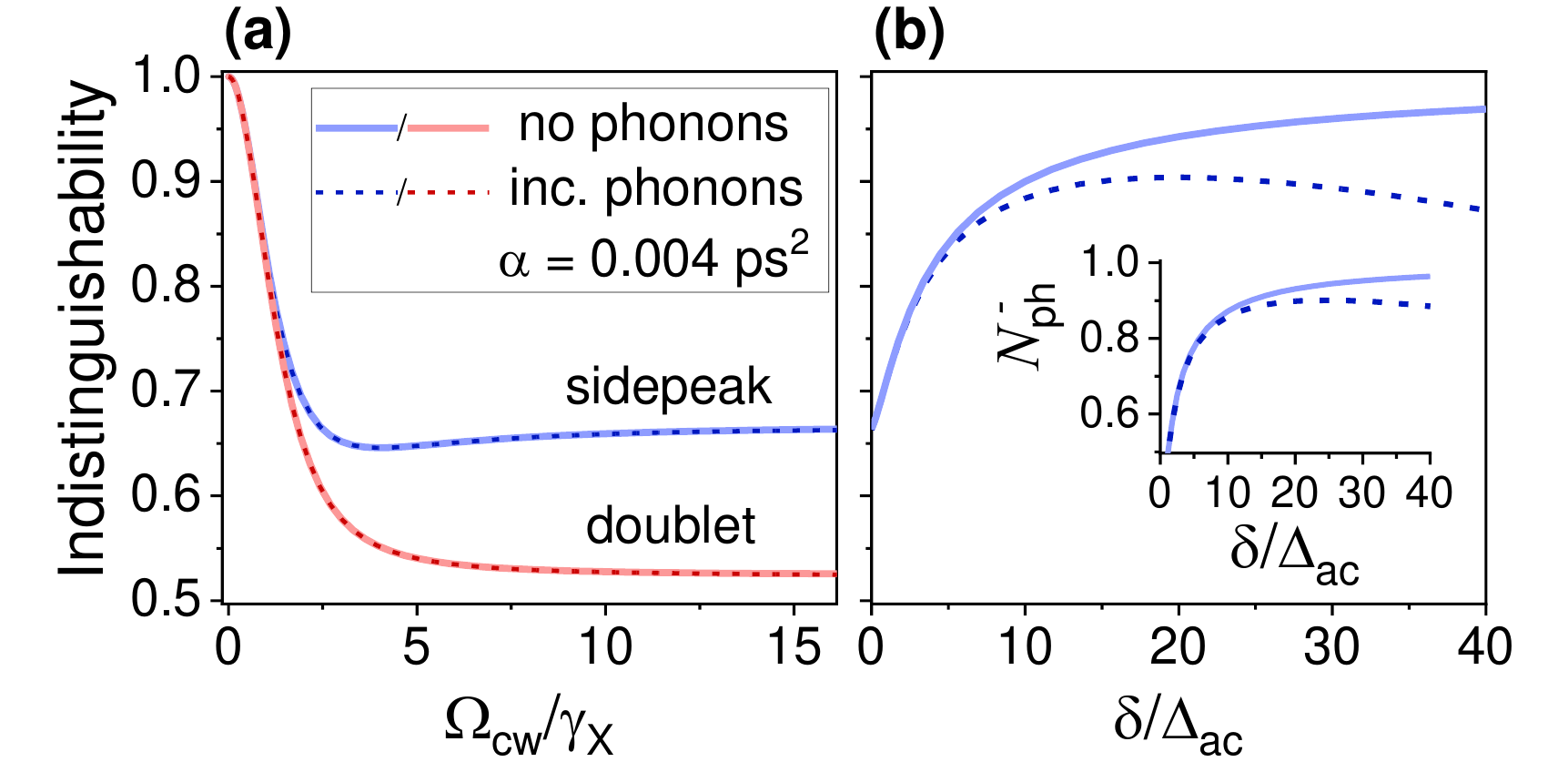}
		\caption{\label{fig:4} {\bf Theoretical expectation of photon indistinguishability.} Simulated indistinguishability $\mathcal{I}$ for the full exciton spectra (red lines) and higher energy exciton peak (blue lines) with (dashed lines) and without (solid lines) electron-phonon coupling
	     for \textbf{(a)} the AT regime ($\delta=0$), and \textbf{(b)} the ac Stark shift regime. In a, we plot $\mathcal{I}$ as a function of dressing strength $\Omega_{\rm cw}$. In b, the cw drive strength is constrained such that for the given laser detuning $\delta$, an ac stark shift of $\hbar \Delta_{\rm ac} = -12 \ \mu$eV is achieved (blue shift). The inset in b shows the $N_{\rm ph}^-$ peak emitted photon number.
		}
	\end{figure}	

    In addition to the observation of AT splitting in the frequency domain, Rabi oscillations should be visible in the time-resolved experiments. To theoretically capture the basic effects of resonant ($\delta{=}0$) cw dressing on the system dynamics, we assume that the short pulse initializes the system in the $\ket{X_V}$ state at $t{=}0$. For strong dressing where $\Omega_{\rm cw}/\gamma_X {\gg} 1$,  neglecting phonons, and assuming the biexciton radiative recombination rate to be two-times larger than that of the V-exciton $\gamma_X$, the $\ket{X_V}$ population $\rho_{X_V}(t)$ is
	\begin{equation}
	    \label{eq:pop}
	    \rho_{X_V}(t) \approx \frac{e^{-\gamma_X t}}{2}\left(1 + e^{-\frac{3}{4}\gamma_X t}\cos{(\Omega_{\rm cw} t)}\right),
	\end{equation}
    yielding damped Rabi oscillations. Analogously, the $\ket{XX}$ population follows similar oscillatory behavior, but anti-phased with respect to $\ket{X_V}$ such that  $\rho_{XX}(t) {=} e^{-\gamma_X t} {-} \rho_{X_V}(t)$. Figure~\ref{fig:3}(e) shows the bare exciton resonance fluorescence decay with a time constant of 500$\pm$10~ps. In Fig.~\ref{fig:3}(f), time-traces of dressed exciton and biexciton emission are shown (30~$\mu$W power cw pump). Both decays exhibit intensity oscillations superimposed on the exponential decay. The observed evolution is direct evidence of Rabi oscillations between the exciton and biexciton with frequency controllable by the dressing laser power (more details in SI Section 6).
  
    Finally, we study how our coherent three-level driving scheme affects the purity, efficiency, and indistinguishability of the emitted single photons. For pump pulses of $\sim$2~ps width, the biexciton-exciton cascade dressing should not have significant effect on the purity of the single-photon emission, given that the timescale of excitation is much shorter than the rate of cw driving, other than the limits set by the two-level pumping re-excitation process~\cite{Hanschke2018,Fischer2018,Gustin2018} as well as a small contribution from the cw off-resonant $\ket{X_V}$ drive (see SI Sections 4, 5).
    
    Contrastingly, the cascade dressing is expected to have an impact on the coherence properties of the exciton emitted photons. In the absence of dressing, the figures of merit for single photons from two-level QD emitters are well-studied~\cite{kiraz04,ilessmith17,Gustin2018,Hanschke2018}. To capture the basic effects of the cw dressing on the indistinguishability of the emitted photons (neglecting phonons and multiphoton emission) we assume the short pulse initializes the system in the $\ket{X_V}$ state. Then, we calculate the $\ket{X_V}$-$\ket{G}$ transition single photon indistinguishability for a Hong-Ou-Mandel (HOM) setup simulated with a Mach-Zehnder interferometer according to~\cite{Loredo2016}
	\begin{equation}
	\mathcal{I} = \frac{2\gamma_X^2}{N_{\rm ph}^2}\int_0^\infty \! dt \int_{0}^\infty \! d\tau |g^{(1)}(t,\tau)|^2,
	\end{equation}
    where $N_{\rm ph} {=} \gamma_X\int dt \langle \sigma_X^+\sigma_X^-\rangle (t)$ is the exciton emitted photon number, and $g^{(1)}(t,\tau){=}\langle \sigma_X^+(t+\tau)\sigma_X^-(t)\rangle$. For well separated emission peaks, we can also define indistinguishabilities for each of the emission peaks $\mathcal{I}^{\pm}$, and decompose the emitted photon number as $N_{\rm ph} {\approx} N^+_{\rm ph} {+} N^-_{\rm ph}$ (see SI Section 4). 
	
    In Fig.~\ref{fig:4}(a), we plot the indistinguishability of emitted photons as a function of dressing strength $\Omega_{\rm cw}$, for both the total doublet spectrum and a single (filtered) emission peak in the AT regime ($|\delta| {\ll} \Omega_{\rm cw}$). For non-zero splittings, the indistinguishability of emitted photons is noticeably reduced which can be attributed to dephasing due to simultaneous dipole radiation from both biexciton and exciton decay channels (known as {\em timing jitter})\cite{Simon2005, Scholl2020}.

    Next, we move from AT to the ac Stark shift regime by increasing both $\delta$ and $\Omega_{\rm cw}$ such that the exciton frequency ac Stark shift of $\hbar \Delta_{\rm ac} {=} -12$ $\mu$eV is kept constant (blue shift), as shown in Fig.~\ref{fig:4}(b). By increasing $\delta/\Delta_{\rm ac}$ from 0 to above 20 indistinguishability is improved from $66\%$ to above $95\%$  and photon emission probability $N_{\rm ph}$ from 50\% to above $90\%$. Such an improvement is attributable to decreased population of the biexciton state during dressing, and shows high potential for this driving scheme.
    
    In solid-state emitters, such as semiconductor QDs, increased $\Omega_{\rm cw}$ results in stronger electron-phonon induced dephasing~\cite{PhysRevX.1.021009,nazir16,Carmele2019}. While the exciton-phonon coupling strength $\alpha$ is not explicitly known in the investigated QD, simulations performed for $\alpha = 0.004 \ \text{ps}^2$ (a value similar to that measured for a similar QD-waveguide setup~\cite{reigue2017}) shows that the highest achievable indistinguishabilities are decreased down to $\sim$90\%.

	\begin{figure}[t]
		\includegraphics[width=0.7\columnwidth]{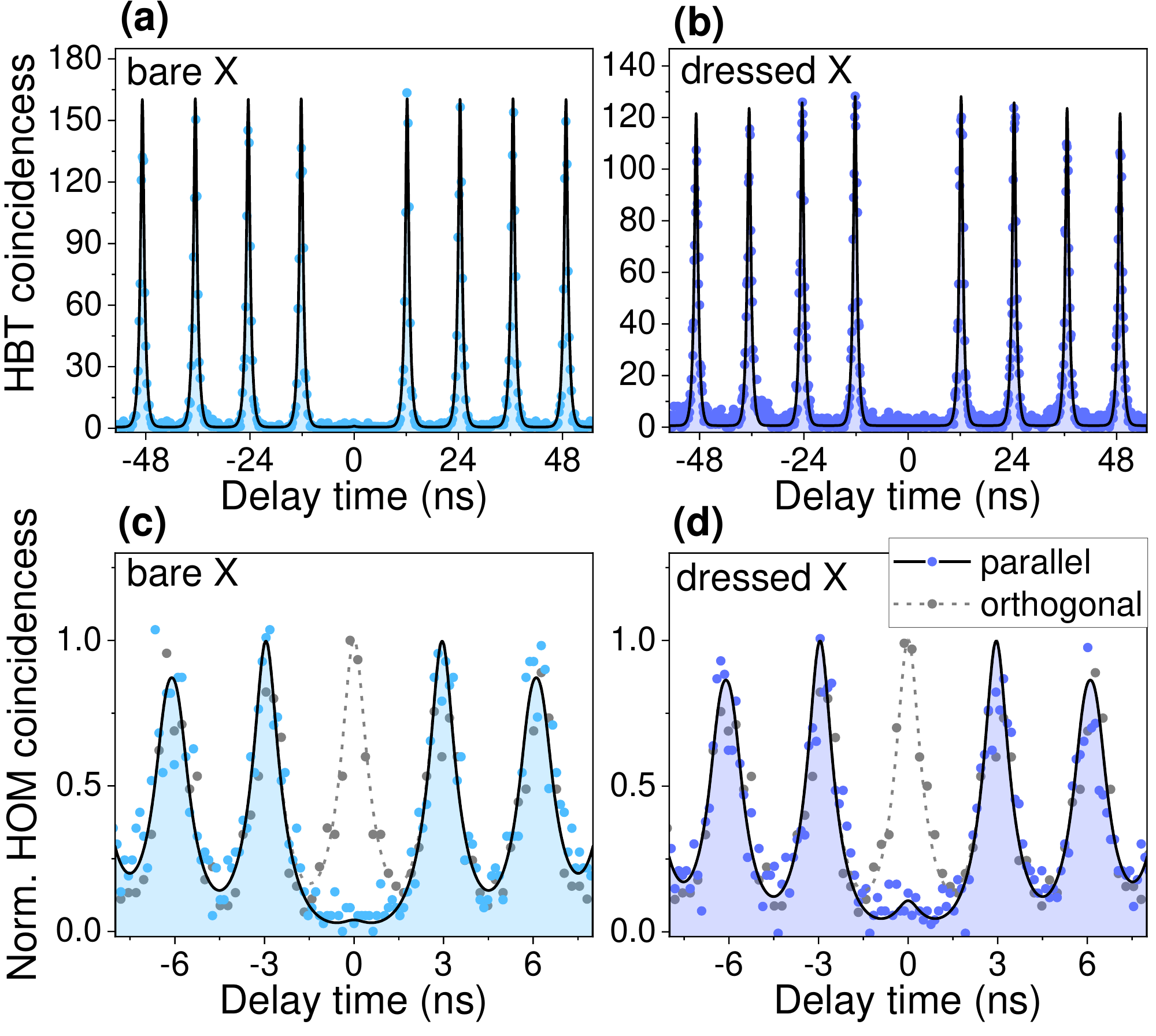}
		\caption{\label{fig:5}\textbf{Indistinguishable single-photon generation.} \textbf{(a)},\textbf{(b)},~Second order correlation histogram of the bare and dressed exciton emission. \textbf{(c)},\textbf{(d)},~Normalized Hong-Ou-Mandel interference histogram of the 3~ns subsequently emitted photons in case of bare and dressed exciton transition. Both HOM and HBT experiments are performed under $\pi$-pulse excitation.
		}
	\end{figure}
	
	Next, we experimentally investigate if indeed high indistinguishability and pure single photon generation is obtainable within our proposed driving scheme. We perform time-correlated measurements on the bare exciton transition serving as a reference, as well as a $\hbar \Delta_{\rm ac} = -12$~$\mu$eV (blue) ac Stark shifted exciton peak, dressed with a cw laser (30~$\mu$W power) slightly detuned from the $|XX\rangle$-$|X_V\rangle$ transition. 
	
	First, we perform second-order correlation experiments in a Hanbury Brown and Twiss (HBT) configuration to check the purity of single-photon emission. Histograms recorded under $\pi$-pulse excitation for exciton emission are presented in Figs.~\ref{fig:5}(a) and b, for the bare and dressed case, respectively. We observe strong suppression of multiphoton emission probability of $g^{(2)}[0]=0.013\pm0.001$ and $g^{(2)}[0]=0.028\pm0.003$. The slight increase of $g^{(2)}[0]$ in the dressed case can be attributed to the presence of the cw background, which we relate to very weak non-resonant excitation of the exciton by cw dressing laser (see SI Section 5). Note that these are essentially upper bounds to the pulse-wise $g^{(2)}[0]$ limited by the detectors dark counts, and our theoretical predictions are consistent with a much smaller multiphoton probability associated with the pulse excitation (see SI Section 4).
	
	To test the photon indistinguishability, HOM interference measurements between two consecutively emitted single photons at a time delay of 3~ns are performed. Figures~\ref{fig:5}(c) and (d) present normalized two-photon interference histograms obtained for photons prepared with parallel (blue points) and orthogonal polarizations (grey points) for the bare and dressed case, respectively. The almost complete lack of the central peak for the co-polarized measurements indicates high indistinguishability of consecutively emitted photons, with two-photon interference visibilities of $\mathcal{V}_{\rm corr}$ = 0.97$\pm$0.03, and $\mathcal{V}_{\rm corr}$=0.94$\pm$0.03 after correcting for the setup interferometer imperfections (see SI). The recorded decrease of the photons indistinguishability due to the dressing agrees well with the theoretical prediction; see SI Section 4 for detailed discussion.
	
	In conclusion, we have demonstrated a new coherent driving scheme of a three-level ladder system by resonant excitation with two laser fields. We have shown the feasibility of this approach towards spectral and temporal tuning of QD-based single-photon sources while allowing for triggered operation with a high degree of photon purity and indistinguishability. Frequency tuning using our technique is especially appealing as it is performed by optical means and thus could be used at very fast speeds allowing for energy control of subsequently emitted photons, giving means for the creation of high-dimensional frequency-encoded quantum states of light~\cite{Shlesinger2021}. While for QDs we expect that exciton-phonon scattering will in general strongly limit the achievable indistinguishabilities of emitted photons for larger energy shifts, this limit can be overcome by the use of, for example, the modified photonic density of states afforded by an optical cavity, which can also be used to bypass the timing jitter which in the case studied here led to low predicted indistinguishability in the AT regime. Our driving scheme could be straightforwardly applied to any other three-level ladder systems, including atoms, ions or vacancy centres in solids. We note that the resonant pulsed drive, while critical to obtain high indistinguishability of on-demand generated single photons, could be replaced by less demanding experimentally excitation techniques such as pumping above-band, through higher energy levels or via two-photon resonances.
	
	The current study was limited to cw dressing of the  $|XX\rangle$-$|X_V\rangle$ transition, resulting in the AT doublet formation (ac Stark shift). It is, however,  possible to modulate the dressing strength in time, giving rise to, largely unexplored for quantum emitters, physics under frequency modulation~\cite{Silveri2017}. This could in turn enable one to spectrally and temporally engineer single-photon states on a time-scale faster than the emitter lifetime, which might allow in our case for temporal and spectral shaping of the single photons, generation of exotic single-photon spectrum beyond the AT doublet, or the formation of topological Floquet states~\cite{Silveri2017,Lukin2020}.	
	
	\begin{acknowledgement}
	The authors thank Silke Kuhn for fabricating the structures. \L{}.D. and Stephen H. acknowledge financial support from the Alexander von Humboldt Foundation. We are furthermore grateful for the support by the State of Bavaria, the Natural Sciences and Engineering Research Council of Canada, and the Canadian Foundation for Innovation.
	\end{acknowledgement}
	
%	\textbf{Author contributions}\\
%	\L{}.D. conceived and designed the experiments. Sven H. and C.S. grew the sample. \L{}.D. designed the device, built the optical set-up, performed the measurements and analyzed the data. C.G. and Stephen H. performed the theoretical modelling. \L{}.D. and C.G. wrote the manuscript with input from all authors. Sven H. provided an experimental infrastructure and managed the project.
	
	\bibliography{bib-manuscript}% Produces the bibliography via BibTeX.content...
	
\end{document}

\title{Supporting Information:\\All-optical Tuning of Indistinguishable Single-Photons Generated in Three-level Quantum Systemss}
	
	\author{\L{}ukasz Dusanowski}
	\email{lukaszd@princeton.edu}
	\affiliation{Technische Physik and W\"{u}rzburg-Dresden Cluster of Excellence ct.qmat, Physikalisches Institut and Wilhelm-Conrad-R\"{o}ntgen-Research Center for Complex Material Systems, University of W\"{u}rzburg, Am Hubland, D-97074 W\"{u}rzburg, Germany}
	\affiliation{Present address: Department of Electrical and Computer Engineering, Princeton University, Princeton, NJ 08544, USA}
	
	\author{Chris Gustin}
	\affiliation{Department of Applied Physics, Stanford University, Stanford, California 94305, USA}
	\affiliation{Department of Physics, Engineering Physics, and Astronomy, Queen's University, Kingston, Ontario K7L 3N6, Canada}
	
	\author{Stephen Hughes}
	\affiliation{Department of Physics, Engineering Physics, and Astronomy, Queen's University, Kingston, Ontario K7L 3N6, Canada}
	
	\author{Christian Schneider}
	\affiliation{Technische Physik and W\"{u}rzburg-Dresden Cluster of Excellence ct.qmat, Physikalisches Institut and Wilhelm-Conrad-R\"{o}ntgen-Research Center for Complex Material Systems, University of W\"{u}rzburg, Am Hubland, D-97074 W\"{u}rzburg, Germany}
	\affiliation{Institute of Physics, University of Oldenburg, D-26129 Oldenburg, Germany}
	
	\author{Sven H\"{o}fling}
	\affiliation{Technische Physik and W\"{u}rzburg-Dresden Cluster of Excellence ct.qmat, Physikalisches Institut and Wilhelm-Conrad-R\"{o}ntgen-Research Center for Complex Material Systems, University of W\"{u}rzburg, Am Hubland, D-97074 W\"{u}rzburg, Germany}
	
	%\date{\today}
	
	\begin{abstract}
	    In this supporting information document,
	    we provide further details on the sample structure (S1), the experimental configurations (S2), characteristics of our quantum dots (S3), the theory used behind results in the main text (S4),  the CW contribution to the pulsed emission signal (S5), time-traces for different dressing field strengths (S6), and the correlation data analysis (S7).
%	    }
	\end{abstract}
	
	%\pacs{85.35.Be, 42.50.Ar, 42.82.-m, 42.50.Dv}
	
	%\keywords{single photon source, Mollow triplet, Autler-Townes effect, resonance fluorescence, quantum dot, two-photon interference}
	
	\maketitle
	%introduction very general regarding two and three-level systems
	
	\renewcommand\thefigure{S\arabic{figure}}   
	
	%sample details
	\section{S1: Sample structure}
	
\begin{figure}[h]
		\includegraphics[width=5.5in]{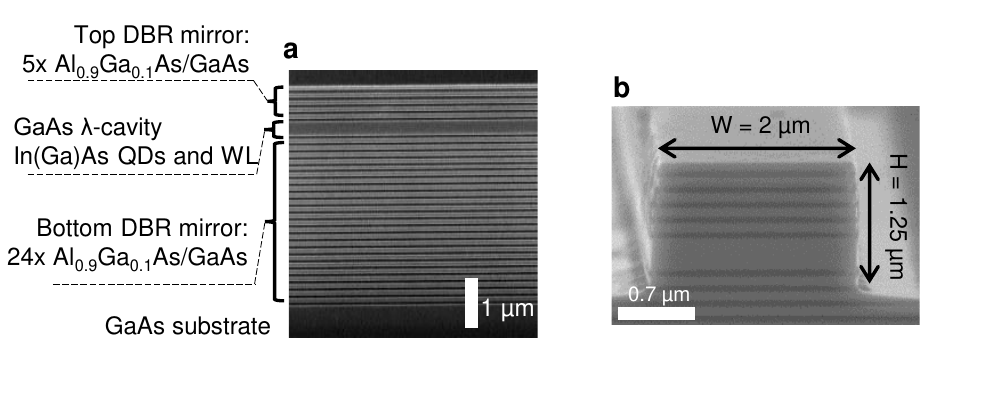}
		\caption{\label{fig:sample}\textbf{Sample structure image.} a,~Scanning electron microscope cross-section image of planar sample with visible layers. The quantum dot (QD) layer is placed inside the centre of a $\lambda$ cavity sandwiched between two distributed Bragg Reflectors consisting of 5/24 alternating $\lambda$/4-thick layers of Al$_{0.9}$Ga$_{0.1}$As and GaAs. b,~Cleaved facet SEM image of the fabricated waveguide device.  
		}
	\end{figure}

	To fabricate our WG device we use a semiconductor sample which contains self-assembled In(Ga)As QDs grown by the Stranski-Krastanow method at the centre of a planar GaAs microcavity. The lower and upper cavity mirrors contain 24 and 5 pairs of Al$_{0.9}$Ga$_{0.1}$As/GaAs $\lambda$/4-layers, respectively, yielding a quality factor of $\sim$200. A $\delta$-doping layer of Si donors with a surface density of roughly $\sim$10$^{10}$~cm$^{-2}$ was grown 10~nm below the layer of QDs to probabilistically dope them. To fabricate ridge WG devices, the top mirror layer along with the GaAs cavity is etched down, forming a ridge with a width of $\sim$2.0~$\mu$m and a height of $\sim$1.25~$\mu$m. Ridges have been defined by e-beam lithography and reactive ion etching. After processing, the sample was cleaved perpendicularly to the WGs. In Fig~\ref{fig:sample}a, a scanning electron microscope (SEM) cross-section image of the planar sample with visible layers is shown. In Fig~\ref{fig:sample}b, the fabricated waveguide device is presented, as viewed from the cleaved sample facet.
	
	\newpage
	\section{S2: Experimental configuration}
	
\begin{figure}[htb]
		\includegraphics[width=5.in]{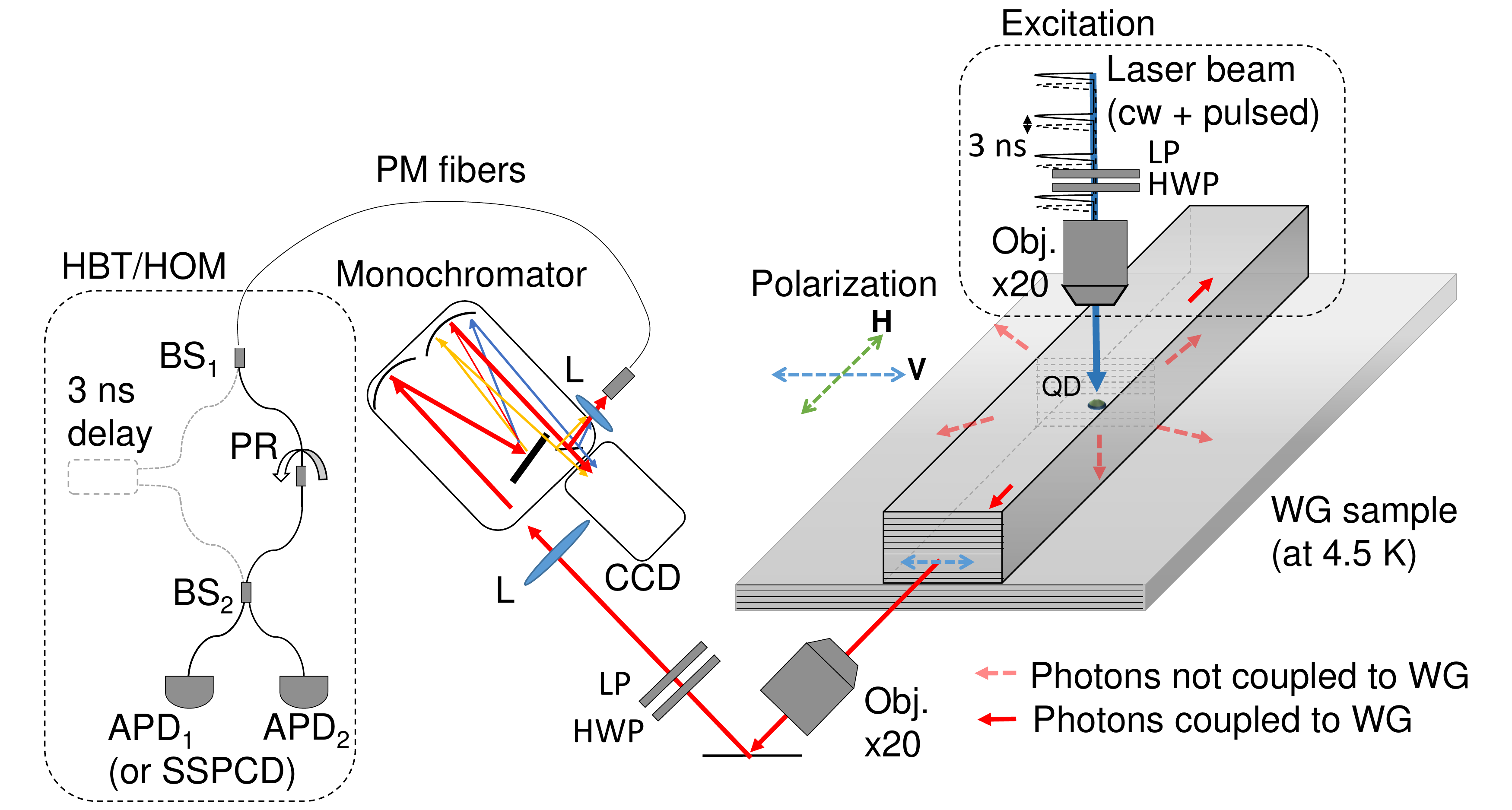}
		\caption{\label{fig:setup} \textbf{Experimental configuration scheme.} Setup used for top excitation (blue path) and side detection (red path) photoluminescence and resonance fluorescence measurements. For polarization control of the excitation laser beam, a half-wave-plate (HWP) combined with a linear polarizer (LP) was used. The polarization orientation is schematically marked with H and V arrows. The side detected signal is analyzed with a 75~cm focal length monochromator combined with a liquid nitrogen cooled CCD detector. For HBT and HOM experiments, a monochromator is used as a spectral filter (75~$\mu$eV bandwith) and the signal is passed into an unbalanced Mach-Zehnder interferometer based on polarization-maintaining (PM) fibres and beam-splitters (BS). In the case of two-photon interference experiments, the QD was excited twice every laser pulse cycle with a delay of 3~ns. For polarization rotation (PR) in the fibre-based HOM interferometer, ceramic sleeve connectors between two fibre facets were used allowing for alignment of the fast and slow axes at the desired angle. For single-photon detection, two avalanche photo-diodes (APD) with 350~ps response times and around 10 dark counts per second were used. For time-resolved measurements, a superconducting single-photon counting detector (SSPCD) was used with a response time of around 30~ps.
		}
	\end{figure}

	For all experiments, the sample is kept in a low-vibration closed-cycle cryostat (attoDRY800) at a temperature of $\sim$4.5~K. The cryostat is equipped with two optical windows allowing for access from both the side and top of the sample. A spectroscopic setup consisting of two independent perpendicularly aligned optical paths is employed, shown schematically in Fig.~\ref{fig:setup}. The QD embedded into the WG is excited from the top through a first microscope objective with NA~=~0.4, while the emission signal is detected from a side facet of the WG with a second objective with the same NA. For non-resonant experiments, we use a cw above-band 660~nm diode laser line. For resonant experiments, we use an external cavity cw tunable diode laser with a linewidth of around 500~kHz and a pulsed Ti:Sa tunable laser with a pulse width of around 2~ps and 82~MHz repetition rate. For fluorescence analysis, the signal is spectrally dispersed by a 75~cm focal length monochromator and focused on a low-noise liquid-nitrogen-cooled CCD camera (around 40~$\mu$eV spectral resolution). For HBT and HOM experiments, the monochromator serves as a spectral filter with 70~$\mu$eV width and signal is introduced into an unbalanced 3-ns delayed Mach-Zehnder interferometer based on single-mode polarization-maintaining optical fibres connected with avalanche single-photon counting detectors (350~ps time-response). For HBT experiments one of the interferometer arms is blocked. For the time-resolved experiments, a fast superconducting single-photon detector with 30~ps resolution is used. A schematic of the experimental setup can be found in SI Section 2.
	
	\newpage
	\section{S3: Basic characteristics of the investigated quantum dot}
\begin{figure}[h]
		\centering
		\includegraphics[width=4.3in]{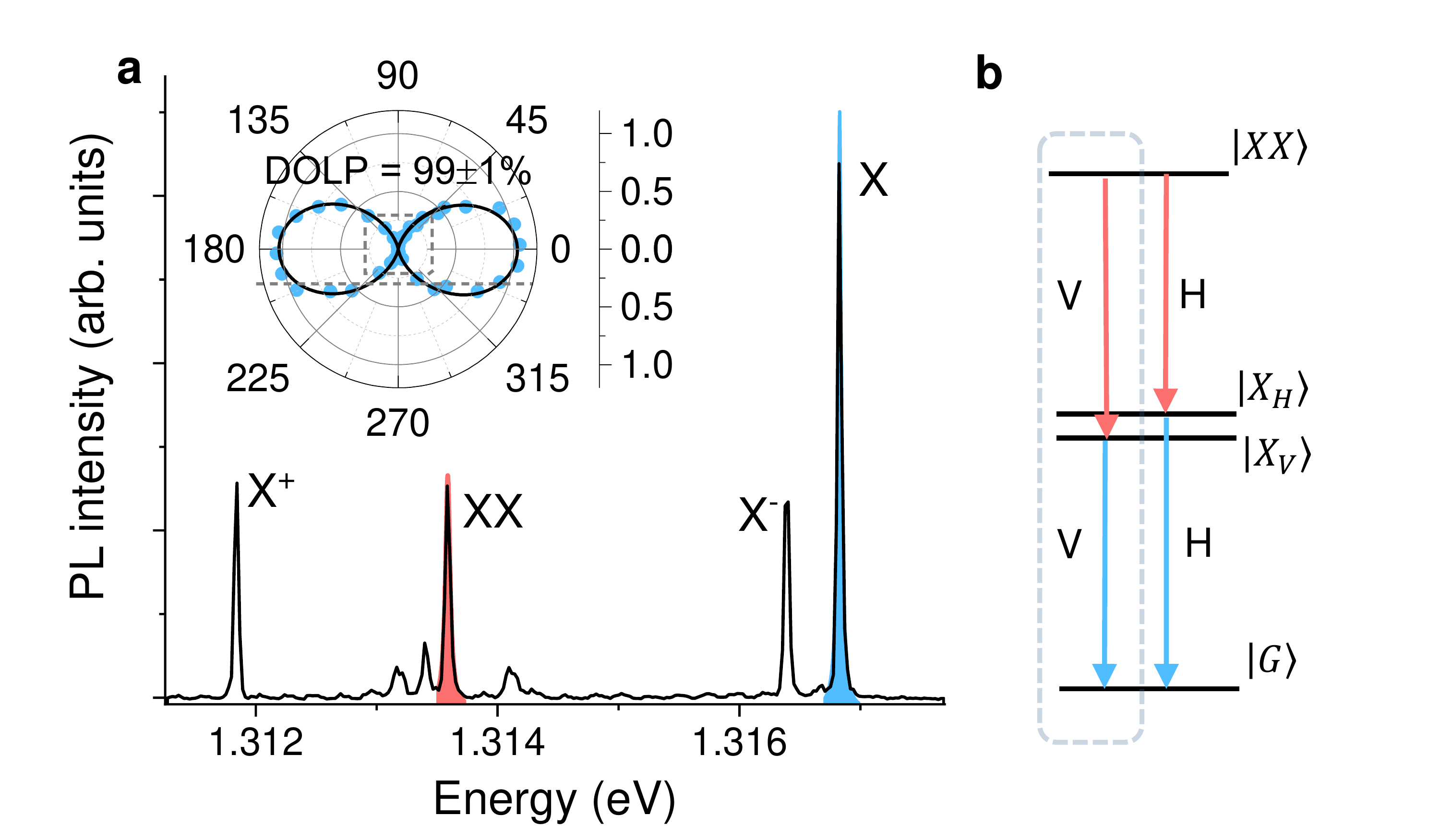}
		\caption{\label{fig:PL}
			\textbf{Investigated quantum dot emission spectra.} a,~Photoluminescence spectrum of the investigated quantum dot under cw above-band excitation. Lines labelled as exciton ($X$), biexciton ($XX$), positively charged exciton ($X+$) and negatively charged exciton ($X-$) are identified, based on power- and polarization-resolved investigations (see \ref{fig:pol}). Inset: polar plot of the exciton line intensity vs detected linear polarization angle. All emission lines exhibit a degree of linear polarization (DOLP) oriented within the TE mode of the waveguide of over 99\%. b,~Schematic of the biexciton-exciton cascade in the QD. Within the dashed box, vertically polarized transitions are marked, which in the case of the investigated device, correspond to the waveguide coupled part of the emission. 	
		}
	\end{figure}

\begin{figure}[htp]
		\centering
		\includegraphics[width=6in]{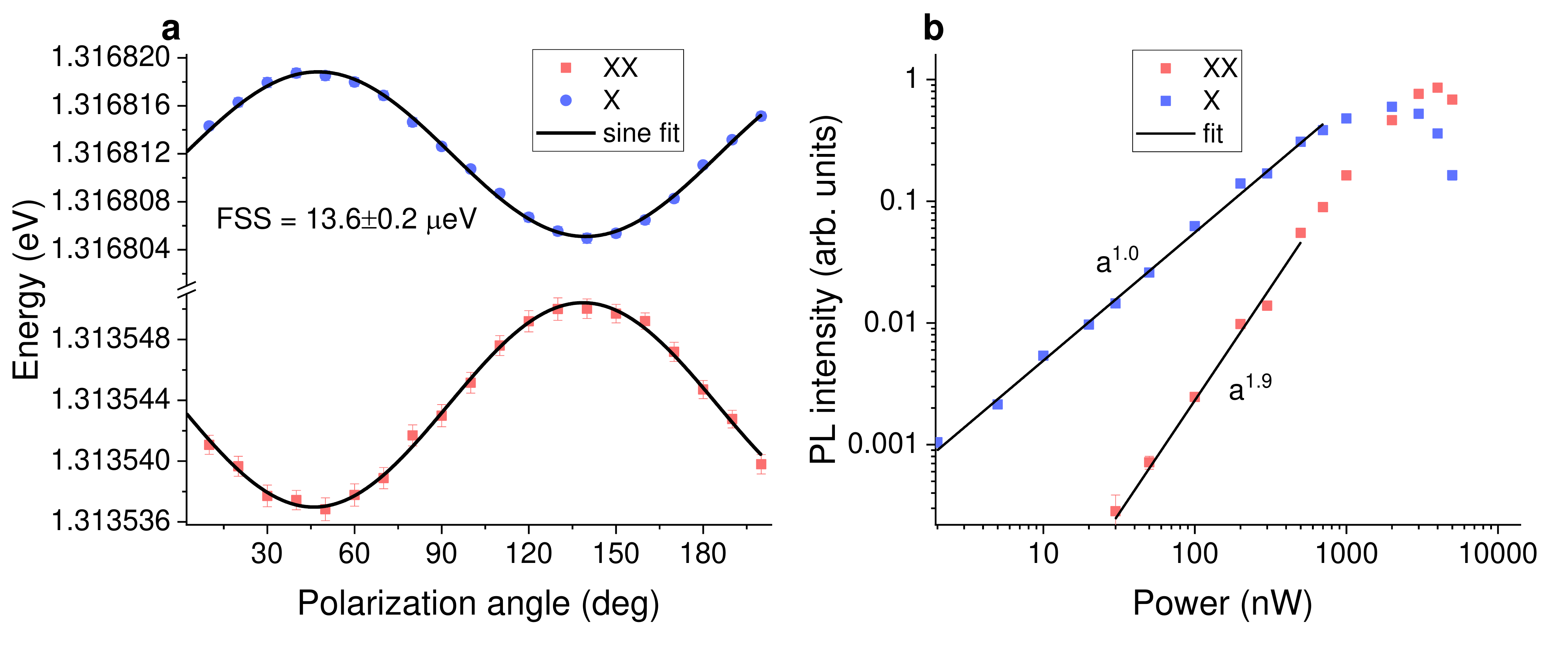}
		\caption{\label{fig:pol}
			\textbf{Polarization and power-resolved photoluminescence.} a,~Emission energy of exciton and biexciton transitions under cw above-band excitation as a function of the detected linear polarization angle, with the signal collected from the top of the waveguide. Clear oscillatory dependence can be observed related to the presence of exciton fine-structure splitting (FSS). Data is fit using the sine function, revealing an exciton FSS of 13.6$\pm$0.2~$\mu$eV. b,~Exciton and biexciton emission intensity vs pumping power presented with a log-log scale. Clear linear and quadratic dependencies can be observed prior to saturation for the exciton and biexciton emissions, respectively.   	
		}
	\end{figure}

	In Fig.~\ref{fig:PL}, a side detected PL spectrum of the QD under investigation is shown. The spectrum is recorded under above-band (660~nm) continuous wave (cw) excitation from the side of the waveguide. A slight modulation of the spectra is visible, related to the Fabry-P\'erot modes present due to reflections between waveguide (WG) facets. Resonance peaks identified as exciton (X), biexciton (XX), positively charged exciton (X+) and negatively charged exciton (X-) are shown, based on power- and polarization-resolved investigations (see \ref{fig:pol}).
	%include polarization-resolved PL
	
	\newpage
	\section{S4: Theory}
	
	To theoretically analyze the dynamics and coherence properties of emitted single photons from the QD dressed ladder system, we use a four-level master equation approach based on the biexciton-exciton cascade, with coherent fields corresponding to the cw and pulse drives. To incorporate electron-phonon scattering, which is important to the semiconductor QD platform, we use a polaron transform master equation. We use the quantum regression theorem to calculate two-time correlation functions, which are needed to compute the emission spectrum and single photon indistinguishability.
	
	In this section, we present our quantum master equation models used to generate the theoretical results of the main text. We first describe the general four-level system model of the dressed biexciton-exciton cascade and its dynamics under incoherent excitation, which we use to compute emission spectra. We then discuss exciton-phonon coupling and the polaron master equation (PME) model we use to incorporate its effects on the dynamics of the driven QD system. We define the figures-of-merit that characterize the single photon source, and finally, we give a comparison of the theoretical prediction for indistinguishability and two-photon interference visibility with experimental data.
	\subsection{Model of biexciton-exciton cascade in QD and dressing of $XX$ and $X_V$ emission under incoherent excitation} 
	
	In this subsection, we consider the system excited incoherently with a far off-resonant cw laser, in addition to the coherent cw laser coupling the biexciton-ground transition. We model the cascade as a four-level system with ground $\ket{G}$, excitons $\ket{X_V}$ and $\ket{X_H}$, and biexciton $\ket{XX}$ states, with the $XX-X_V$ transition dressed by a coherent drive with Rabi frequency $\Omega_{\rm cw}$. For very strong (near resonant) driving, the coherent laser also couples the $X_V-G$ transition. The total Hamiltonian for this setup,  neglecting phonon coupling and radiative emission, is
	\begin{equation}\label{eq:H1}
	    H_{\rm tot} = \hbar \omega_{XX} \sigma^+_{XX}\sigma^-_{XX} + \hbar \omega_{X_H} \ket{X_H}\bra{X_H} + \hbar\omega_{X_V} \sigma^+_{X} \sigma^-_X + \hbar\Omega_{\rm cw}\cos{(\omega_{\rm cw} t)}(\sigma_x^{XX} + \sigma_x^X),
	\end{equation}
	where $\hbar\omega_i$ is the energy of the $i^{\rm th}$ state, $\sigma^+_{XX} = \ket{XX}\bra{X_V}$, $\sigma^-_{XX} = \ket{X_V}\bra{XX}$, $\sigma^+_{X} = \ket{X_V}\bra{G}$, $\sigma^-_X = \ket{G}\bra{X_V}$, $\sigma_x^{XX}=\sigma^+_{XX} + \sigma^-_{XX}$, and $\sigma_x^X = \sigma^+_X + \sigma^-_X$. The undressed system ($\Omega_{\rm cw} =0$) gives rise to $V$-polarized fluorescence emission energies at $E_{XX} = \hbar\omega_{XX} -\hbar\omega_{X_V}$ and $E_X = \hbar\omega_{X_V}$.

	The system is driven at (near)-resonant cw laser frequency $\omega_{\rm cw} = \omega_{XX} - \omega_{X_V} - \delta$,  such that $\delta$ is the laser detuning from the biexciton--V-polarized exciton  transition. By moving into an interaction picture defined by $H_0 = (E_B + \hbar \omega_{\rm cw})\sigma^+_X\sigma^-_X + \hbar\omega_{X_H} \ket{X_H}\bra{X_H} + (E_B + 2\hbar \omega_{\rm cw})\sigma_{XX}^+ \sigma_{XX}^- + E_B\sigma^-_X\sigma^+_X$, where the biexciton binding energy is defined as  $E_B = \hbar(2\omega_{X_V}-\omega_{XX})$, and performing the rotating wave approximation, we obtain the time-independent system Hamiltonian:
	\begin{equation}\label{eq:r1}
	H_S = 2\hbar\delta\sigma^+_{XX}\sigma^-_{XX} +\hbar\delta\sigma^+_X\sigma^-_X - E_B\sigma_X^- \sigma_X^+ +  \frac{\hbar\Omega_{\rm cw}}{2}
	\left(\sigma_x^{XX}+\sigma_x^X\right).
	\end{equation}
	
	Without yet considering phonon coupling, we can model the radiative emission and incoherent excitation using a Lindblad master equation for the reduced density operator of the system $\rho$:
	\begin{align}\label{eq:me1}
	\dot{\rho} =& -\frac{i}{\hbar}[H_S,\rho] + \mathbb{L}_{\text{rad}}\rho + \mathbb{L}_{\text{deph}}\rho + \frac{\gamma_{\rm{inc}}}{2}\mathcal{L}\Big[\ket{XX}\bra{G}\Big]\rho,
	\end{align}
	where we have included radiative decay from both excitons with rate $\gamma_X$, and from the biexciton with \emph{total} rate $\gamma_{XX}$ (i.e., we assume throughout orthogonal polarization channels have equal decay rates):
	\begin{equation}
	\mathbb{L}_{\text{rad}}\rho=\frac{\gamma_X}{2}\mathcal{L}[\sigma^-_X]\rho + \frac{\gamma_X}{2}\mathcal{L}\big[\ket{G}\bra{X_H}\big]\rho+ \frac{\gamma_{XX}}{4}\mathcal{L}[\sigma^-_{XX}]\rho + \frac{\gamma_{XX}}{4}\mathcal{L}\big[\ket{X_H}\bra{XX}\big]\rho.
	\end{equation}
	
	We also include pure dephasing (which can act as phenomenological spectral linewidth broadening) for each transition, such that
	\begin{equation}
	\mathbb{L}_{\text{deph}}\rho = \frac{\gamma'}{2}\mathcal{L}[\sigma^+_X\sigma^-_X]\rho + \frac{\gamma'}{2}\mathcal{L}\Big[\ket{X_H}\bra{X_H}\Big]\rho + \gamma'\mathcal{L}[\sigma^+_{XX}\sigma^-_{XX}]\rho,
	\end{equation}
	where $\mathcal{L}[A]\rho = 2A\rho A^{\dagger} - A^{\dagger}A\rho - \rho A^{\dagger}A$ is the Lindblad operator, and we have included incoherent excitation of the biexciton with rate $\gamma_{\text{inc}}$. The V-polarized incoherent emission spectrum is $S_{\rm{inc}}(\omega)  = S_{X_V}(\omega) + S_{XX}(\omega)$, where
	\begin{equation}\label{eq:xsp}
	S_{X_V}(\omega) = \rm{Re}\Big[ \!\int_0^{\infty} \! \!dt\!\!\int_0^\infty \!\! d\tau  e^{i\big(\omega-(\tilde{\omega}_{X_{V0}}-\tilde{\omega}_{G_0})\big)\tau} \langle C_X^{\dagger}(t)C_X(t+\tau)\rangle \Big],
	\end{equation}
	where $C_X(t) = \sigma^{-}_X(t) - \langle \sigma^{-}_X\rangle(t)$, and 
	\begin{equation}
	S_{XX}(\omega) = \rm{Re}\Big[ \!\int_0^{\infty} \! \!dt\!\!\int_0^\infty \!\! d\tau  e^{i\big(\omega-(\tilde{\omega}_{XX_0}-\tilde{\omega}_{X_{V0}})\big)\tau} \langle C_{XX}^{\dagger}(t)C_{XX}(t+\tau)\rangle \Big],
	\end{equation}
	with $C_{XX}(t) = \sigma^{-}_{XX}(t) - \langle \sigma^{-}_{XX}\rangle(t)$~\cite{cui06}. The frequency $\tilde{\omega}_{i_0}$ is the frequency of the $i^{\rm th}$ energy level that appears in the $H_0$ Hamiltonian used to transform into a rotating frame; as we use multiple rotating frames throughout this analysis, we have left this generic here. For the case of incoherent excitation, the system has a nontrivial stationary state and the $t$-integral can be replaced with a limit $t \rightarrow \infty$. We calculate the spectrum and other two-time correlation functions with the aid of the quantum regression theorem.
	
	For $E_B \gg |\hbar\delta|,\hbar\Omega_{\rm cw}$, we can neglect the $\sigma_x^X$ term in the Hamiltonian, as the exciton-ground transition is driven far off resonance. It is then useful to move into a different moving frame, giving the system Hamiltonian \begin{equation}\label{eq:r2}
	H_S = \frac{\hbar\delta}{2}\sigma_z^{XX} + \frac{\hbar\Omega_{\rm cw}}{2}\sigma_x^{XX},
	\end{equation} 
	where $\sigma_z^{XX} = \sigma^+_{XX}\sigma^-_{XX} - \sigma^+_{X}\sigma^-_{X}$, and the eigenvalues and eigenstates are given in the main text. 
	For the results in the main text, we use in our simulations parameters that match experimental values --- namely, $E_B = 3.24 \ \text{meV}$, $\hbar\gamma_X = 1.32 \ \mu$eV, and $\hbar\gamma_{XX} = 2.53 \ \mu$eV (see Section~S6). For Fig.~2 in the main text, We use a dephasing rate which acts primarily to broaden the spectral linewidths of $\hbar\gamma' = 20 \ \mu$eV, and use a small incoherent excitation rate to ensure the spectra remain in the linear excitation regime. We also include exciton-phonon coupling, discussed in the next section. The deviation in predicted spectral lines from the dressed state model and the full simulation is an indication of exciton-ground dressing for high pump strengths. To investigate the role of this additional dressing on the system energy levels, we can adiabatically eliminate the state $\ket{G}$ in Eq.~\eqref{eq:r1} and then moving into the same rotating frame as Eq.~\eqref{eq:r2}; we find new eigenenergies $E_{\pm}' = \frac{\hbar^2\Omega_{\rm cw}^2}{8E_B} \pm \frac{\hbar}{2}\sqrt{\frac{\Omega_{\rm cw}^4}{16E_B^2} + \eta^2 - \frac{\delta \Omega_{\rm cw}^2}{2E_B}}$, from which the correction to the spectral line positions can be readily obtained. In particular, for resonant excitation with $\delta=0$ we find $E_{\pm}' = \pm \frac{\hbar\Omega_{\rm cw}}{2} + \frac{\hbar^2 \Omega_{\rm cw}^2}{8E_B} + \mathcal{O}\left(\frac{\hbar^3\Omega_{\rm cw}^3}{E_B^2}\right)$, and so to lowest order in the corrections the position of the spectral lines in the biexcitonic Mollow triplet is unaffected, allowing for a simple measurement of $\Omega_{\rm cw}$ by the position of spectral resonances as a function of laser power (for sufficiently low drive strengths, a measurement of the exciton splitting also suffices). 
	
	\subsection{Decomposition of output photon observables under strong splitting}
	For strong driving, as the dressing laser splits the exciton emission spectrum into a doublet with two peaks corresponding to emission from the $\ket{\pm}$ states, it is convenient to
	separate  the emitted photon number into parts arising from each peak. To do so, we decompose the emitted photon number using Eq. (1) of the main text to express the exciton creation and destruction operators in terms of the dressed state basis:
	\begin{align}\label{eq:npm}
	    N_{\rm ph} &= \gamma_X \int_0^{\infty} dt \langle \sigma^+_{X}\sigma^-_{X}\rangle  \nonumber \\ & = \frac{\gamma_X}{2}\left (1-\frac{\delta}{\eta}\right)\int_0^{\infty} dt \rho_+(t) + \frac{\gamma_X}{2}\left (1+\frac{\delta}{\eta}\right)\int_0^{\infty} dt \rho_-(t)  - \frac{\gamma_X \Omega_{\rm cw}}{2\eta}\int_0^{\infty} dt \text{Re}\big\{\rho_{+-}(t)\big\} \nonumber \\ & \approx N^+_{\rm ph} + N^-_{\rm ph},
	\end{align}
	where $\rho_{\pm}(t) = \bra{\pm} \rho(t) \ket{\pm}$, and $\rho_{+-}(t) = \bra{+}\rho(t)\ket{-}$. In the last line, we have dropped an integration over a coherence term, as in the limit of well-separated peaks (i.e., with center-frequencies separated by $\gg \gamma_X$), the integrand is highly oscillatory and contributes negligibly.
	
	To better appreciate why this decomposition corresponds to separate frequency components in the emission spectrum, recall that in a quantum optical input-output theory, the collapse operator $c$ in the Lindblad equation for spontaneous emission $\mathcal{L}[c]\rho$ is proportional to the (scattered) outgoing electric field photon destruction operator~\cite{Gardiner1985}. Using Eq.~(2) in the main text to express $\sigma^-_{X}$ as a superposition of $\ket{G}\bra{\pm}$ operators (which oscillate in the Heisenberg picture with frequency corresponding to their respective spectral shifts), we find in the master equation a sum of spontaneous emission terms proportional to $\mathcal{L}[\ket{G}\bra{\pm}]\rho$, as well as a highly oscillatory term which again averages out in the limit of well-separated peaks.
	
	\subsection{Exciton-phonon coupling} 
	We also incorporate in our calculations the effects of microscopic coupling of the excitons to longitudinal acoustic phonons, which are well-known to have significant effects on the dynamics and spectra of QD excitons~\cite{roy11,nazir16,PhysRevB.80.201311,ramsay10,denning20,Forstner2003}. Some important phonon-scattering effects include the formation of a broad phonon sideband in the exciton emission spectrum~\cite{ilessmith17}, coherent drive-induced decoherence (including dephasing)~\cite{Forstner2003,mccutcheon10}, and phonon-assisted excitation in the presence of a detuned coherent drive~\cite{Gustin2019,cosacchi19}. In this work, we study the effects of phonons using a perturbative polaron master equation (PME) with a Born-Markov approximation. The PME involves a unitary transformation to a ``polaron" frame in which the linear exciton-phonon interaction is exactly solved via the independent boson model in the absence of pump drives. The presence of a cw or pulse laser drive then gives a perturbation Hamiltonian involving polaron-phonon couplings which can be treated using usual (Markovian) master equation methods. For our calculations, we adapt the (multi-excitonic, pulse-dependent) model from Ref.~\cite{Gustin2018}. We use the phonon spectral density $J(\omega) = \alpha \omega^3 \exp{\left[-\frac{\omega^2}{2\omega_b^2}\right]}$, with $\alpha = 0.004 \ \text{ps}^2$, and $\omega_b = 8.36 \ \text{ps}^{-1}$ ($\hbar\omega_b = 5.5 \ \text{meV}$)--- parameters which reproduce our data well and are similar to those extracted from other experiments in waveguides~\cite{reigue2017}. We note that in the regimes studied here (which correspond to a filtered zero-phonon line), once we have accounted for the coherent renormalization of the Rabi drives, the effect of phonons is largely unaffected by the value of $\omega_b$. We also use a temperature $T = 4.5$ K, as measured in experiments.
	
    The PME gives a constant small polaron shift of exciton and biexciton resonances which we absorb into the initial definition of the energy levels. The phonon interaction also causes a coherent renormalization of the effective Rabi drive for pulse and cw lasers, attenuating the effective drive strength (for our phonon parameters and temperature, the amplitudes are reduced by a factor of $\langle B \rangle \approx 0.87$)~\cite{ilessmith17}; we define our pulsed and cw drive strengths $\Omega$ as the \emph{post}-attenuation values. The broad phonon sideband in the emission spectrum that is given analytically by the PME method is neglected here, as this is filtered out in our experiments. This filtering process would reduce the overall efficiency of the source by an additional factor of $\langle B \rangle^2 \approx 0.75$, for our phonon parameters and temperature, which we have not accounted for explicitly~\cite{ilessmith17}.
	
	\subsection{Pulsed resonance fluorescence of $X_V$ under $X_V$-$XX$ dressing}
	
	For calculations where we are including the effect of the excitation pulse, we use the system Hamiltonian
	\begin{equation}\label{eq:H2}
	H_S(t) = \frac{\hbar\Omega_{\rm cw}}{2}\sigma_x^{XX} + \frac{\hbar\Omega(t)}{2}\sigma_x^X,
	\end{equation}
	corresponding to resonant pulsed excitation and resonant cw drive ($\delta=0$), neglecting the far off-resonant cw driving of the $G-X_V$ transition. We use a pulse $\Omega(t)=\Omega_0\exp{\left[-\frac{(t-t_0)^2}{\tau_p^2}\right]}$, with area $\int_{-\infty}^{\infty} dt \Omega(t) = \pi$ such that $\Omega_0\tau_p = \sqrt{\pi}$. The laser pulse width in \emph{intensity} has full width at half maximum 2~ps, which corresponds to $\tau_p = 1.7$ ps. Since $\tau_p \omega_b \gg 1$, the pulse evolves slowly relative to the phonon bath relaxation timescale, and we make the ``additional Markov approximation'' in the PME to treat the time-dependence of the pulse (see Ref.~\cite{Gustin2018}.) We model the pulse explicitly only for Fig. 3(a) and Fig. 3(d) in the main text, as well as $g^{(2)}[0]$ calculations; the other results simply assume an initially excited $X_V$ state (or the ground state $G$, for the spectra calculations).
	
	\subsection{Single photon efficiency, purity, and indistinguishability under pulsed excitation with $X_V$-$XX$ dressing}
	A single photon source is primarily characterized by three figures-of-merit: the efficiency (or brightness), purity (related to the chance of emitting more than one photon), and the indistinguishability, which quantifies the wave packet overlap (or trace purity) of successively emitted single photons.
	To characterize the efficiency of the source, we use the emitted photon number $N_{\rm ph}$ (and $N_{\rm ph}^{\pm}$), defined in Eq.~\eqref{eq:npm}.

	We define the $g^{(2)}[0]$ metric which quantifies the purity of photons emitted from the $X_V-G$ transition in the usual way:
	\begin{equation}
	    g^{(2)}[0] = \frac{2\gamma_X^2}{N_{X_V}^2} \int_0^\infty \! dt \int_{0}^\infty \! d\tau g^{(2)}(t,\tau),
	\end{equation}
	where $g^{(2)}(t,\tau) = \langle \sigma^+_X(t)\sigma^+_X(t+\tau)\sigma^-_X(t+\tau)\sigma^-_X(t)\rangle$. This definition is consistent with that measured in our HBT experiment.	Furthermore, we can obtain this quantity for photons emitted into a single spectral peak (which corresponds experimentally to filtering out the other peak):
	\begin{equation}\label{eq:ipm}
	    g^{(2)}_{\pm}[0] =  \frac{\int_0^{\infty} dt \int_0^{\infty} d\tau g_{\pm}^{(2)}(t,\tau)}{\frac{1}{2}\left[\int_0^{\infty} dt \rho_{\pm}(t)\right]^2},
	\end{equation}
	where $g^{(2)}_{\pm}(t,\tau) =\langle \sigma^+_{\pm}(t)\sigma^+_{\pm}(t+\tau)\sigma_{\pm}^-(t+\tau)\sigma^-_{\pm}(t)\rangle$, $\sigma^-_{\pm} = \ket{G}\bra{\pm}$, $\sigma^+_{\pm} = \ket{\pm}\bra{G}$, and $\rho_{\pm} = \bra{\pm} \rho\ket{\pm}$. In the ac Stark regime, one of these peaks becomes dominant and its characteristics become very similar to that of the total spectrum.
	
	For pulses of intensity FWHM of $\sim \! 2$ ps, we do not expect the cw laser dressing to have a highly significant effect on the multiphoton emission probability of our single-photon source, given that the timescale of excitation is much shorter than the rate of cw driving and spontaneous emission. The two-photon statistics, in this case, can be expected to roughly follow similar trends as in several studies on pulse re-excitation, for example, Refs.~\cite{Hanschke2018,Fischer2018,Gustin2018}. However, the dressing can lower the efficiency of the photon emission process in general, which can increase $g^{(2)}[0]$. Carrying out full simulations including the pulse, we find in both AT and ac Stark regimes values of $g^{(2)}[0]$ and $g^{(2)}_{\pm}[0]$ on the order of $\sim 10^{-3}$, similar to the undressed excitation case (without considering phonons), where we have $g^{(2)}[0] = 1.75 \times 10^{-3}$, in agreement with the analytical result derived by Fischer \emph{et al.}~\cite{Fischer2018}. In the ac Stark regime in particular (where $N_{\rm ph}$ is near-unity), the $g^{(2)}[0]$ of the source asymptotically approaches this value. This number is not noticeably changed in the presence of phonon coupling with our phonon parameters.
	
	We neglect here a small contribution due to cw excitation discussed in Section~S5, as our definition of $g^{(2)}[0] $ is defined for pulsed excitation; we assume this contribution can be accounted for in experimental data to a good approximation by fitting the HOM coincidence counts with a constant offset, given the very fast excitation rate ($\sim \eta$) relative to the system relaxation rate ($\sim \gamma_X$) in the ac Stark regime.
	
    We can calculate the $X_V-G$ transition single photon indistinguishability according to
	\begin{equation}
	\mathcal{I} = \frac{2\gamma_X^2}{N_{\rm ph}^2}\int_0^\infty \! dt  \int_{0}^\infty \! d\tau |g^{(1)}(t,\tau)|^2,
	\end{equation}
	where $g^{(1)}(t,\tau)=\langle \sigma^+_X(t+\tau)\sigma^-_X(t)\rangle$. This definition goes to zero for perfectly distinguishable \emph{single} photons (the high dephasing limit), tends to $1$ for indistinguishable single photons, and is equal to the visibility of two-photon interference observed in a perfect Mach-Zehnder interferometer with zero source multiphoton probability ($g^{(2)}[0]=0$). Furthermore, we can obtain the indistinguishability of photons emitted into each peak:
	\begin{equation}\label{eq:ipm}
	    \mathcal{I}^{\pm} =  \frac{\int_0^{\infty} dt \int_0^{\infty} d\tau |g^{(1)}(t,\tau)|^2}{\frac{1}{2}\left[\int_0^{\infty} dt \rho_{\pm}(t)\right]^2},
	\end{equation}
	where $g^{(1)}_{\pm}(t,\tau) =\langle \sigma^+_{\pm}(t+\tau)\sigma^-_{\pm}(t)\rangle$.

    \subsection{Comparison with two-photon interference measurements}
    The raw two-photon interference visibility as measured in our HOM experiment can be expressed in terms of the above quantities as~\cite{Santori2002}
    \begin{equation}
        \mathcal{V}_{\rm raw} = \frac{\mathcal{I}_{X_V}/\chi_{\rm corr} - g^{(2)}[0]}{1 + g^{(2)}[0]},
    \end{equation}
    where 
    \begin{equation}
        \chi_{\rm corr} = \frac{R^2+T^2}{2RT(1-\epsilon)^2},
    \end{equation}
    is a correction factor related to imperfections associated with the interferometry setup; $(1-\epsilon) = 0.99$ is the interferometer fringe contrast, and $R/T = 1.15$ is the beam-splitting reflectance to transmittance ratio, leading to $\chi_{\rm corr} \approx 1.0303$.

    For the ac Stark regime $\delta_{\rm ac} \gg \Delta_{\rm ac}$, where the exciton-emitted spectrum takes on a single frequency-shifted peak, the results of the previous section suggest that the $g^{(2)}[0]$ is over an order of magnitude smaller than the experimental uncertainty associated with $\mathcal{V}_{\rm raw}$, and can be neglected. Thus, we can compare this experimental measurement with the indistinguishability simulations used in Fig. 4 of the main text, where the system is initialized in the $X_V$ state and consequently $g^{(2)}[0] =0$, by the relation 
    \begin{equation}
     \mathcal{V}_{\rm corr} \approx \chi_{\rm corr} \mathcal{V}_{\rm raw} + \mathcal{O}(g^{(2)}[0]),
    \end{equation}
    and $\mathcal{V}_{\rm corr} \approx \mathcal{I} + \mathcal{O}(g^{(2)}[0])$.

    For the undressed (bare exciton) case, we have $\mathcal{V}_{\rm corr} = 0.97 \pm 0.03$, which is consistent with (near)-unity indistinguishability as predicted by our theoretical model, and for the dressed case, we have $\mathcal{V}_{\rm corr} = 0.94 \pm 0.03$ for an ac Stark shift of $\hbar \Delta_{\rm ac} = -12 \ \mu \text{eV}$. This result is in reasonable agreement with our theoretical prediction using $\alpha = 0.004 \ \text{ps}^2$ for the phonon coupling strength, where we find a maximum indistinguishability of $\mathcal{I} \approx 0.904$ which occurs at $\delta_{\rm ac} \approx 20 \Delta_{\rm ac}$. We note that we have made no specific attempt to fit our phonon coupling parameters to experimental data, and of course greater agreement with the experimental result could be obtained by simply using a lower value of phonon coupling $\alpha$.

    We additionally note that as a consequence of the analysis in this section, we estimate that the upper-bound to the $g^{(2)}[0]$ as reported in the main text is likely a significant overestimate of the actual pulse-wise value.
	\newpage
	\section{S5: Cw contribution into pulsed emission signal} \label{sec:S6}
	
\begin{figure}[hbt]
		\centering
		\includegraphics[width=6.0in]{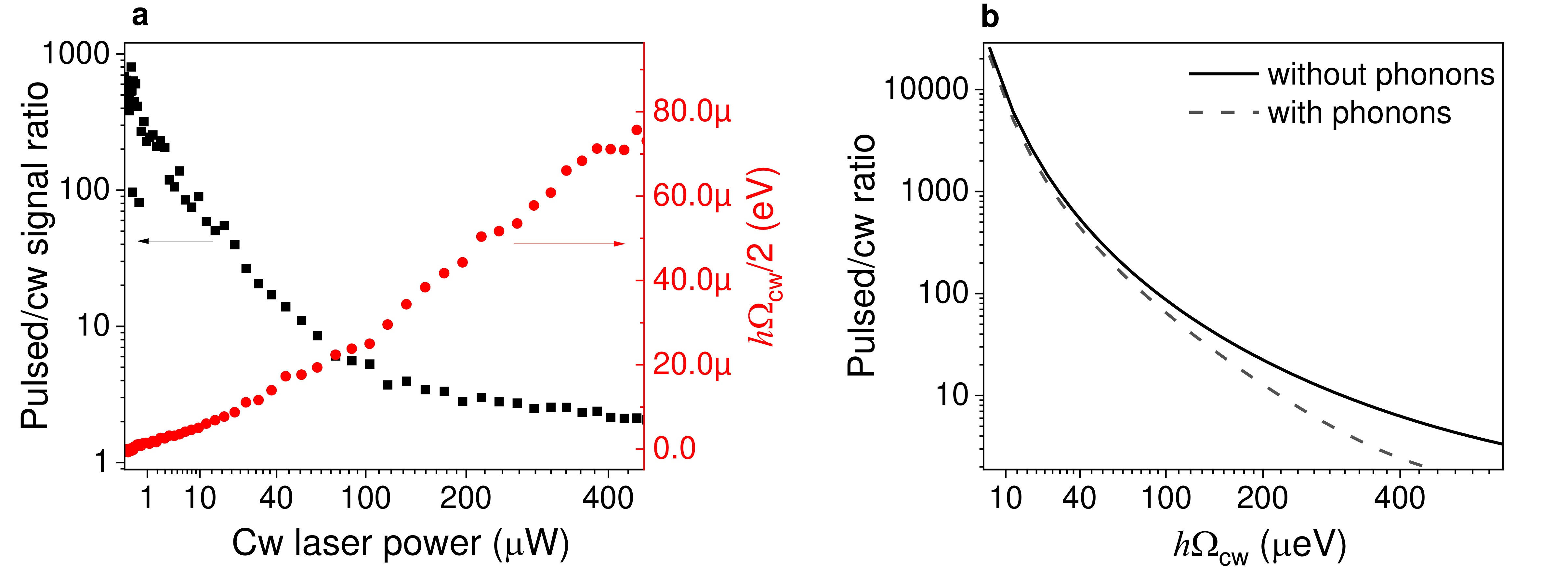}
		\caption{\label{fig:ratio}
			\textbf{Non-resonant vs resonant excitation contribution to $X_V$ emission} a,~Ratio of pulsed vs cw signal contribution in exciton emission as a function of the cw driving laser power resonant with the $XX-X_V$ transition. To estimate the cw contribution to the emission signal, spectra under cw+pulse drive and cw drive only were recorded, respectively. b,~Simulated ratio of pulsed signal in the presence of cw dressing to signal in presence of just cw dressing.
		}
	\end{figure}

	In our experiment, pulsed excitation of the exciton state leads to the triggered emission of photons synchronized with the laser pump via resonance fluorescence. The presence of the cw drive dressing the exciton-biexciton resonance additionally raises the potential of undesirable excitation of the system from its ground state due to the far-off resonance driving of the ground-exciton transition, as is present in Eq.~\eqref{eq:H1}. In our proposed driving scheme, the cw excitation of the $X_V$ state should be minimized as the dressing laser energy is much below the energy of the exciton resonance. To quantify this effect, exciton emission spectra as a function of the cw drive power was recorded under cw + pulse drive and cw drive only, respectively. In Fig.~\ref{fig:ratio}a, the experimental pulsed to cw emission intensity ratio as a function of cw excitation power is shown. It is important to note here that the recorded ratio corresponds to a cw drive of $XX$ and a pulsed drive of $X_V$ with a repetition rate of 82~MHz, and could be potentially improved by increasing the pulsed laser repetition rate and modulation of the cw drive. In Fig.~\ref{fig:ratio}b we show the theoretically simulated ratio of pulsed signal in the presence of cw dressing to signal in presence of just cw dressing. To generate this plot, we assume for the ``pulse'' case an initial condition of $\ket{X_V}$, as this is an excellent approximation for short pulse excitation, and calculate the cumulative emitted photon number $N_{\rm ph}$ by a time $t = [82 \text{MHz}]^{-1} = 12.2$ ns, then normalize this value to its value for the same simulation with initial value $\ket{G}$. For $E_B=3.24$ meV, at large drive strengths phonons play a role in increasing the cw excitation of $X_V$, due to phonon absorption-assisted excitation. This effect is relatively weak for small drive strengths, which can be understood as a consequence of the low thermal occupation of acoustic phonons with energy $E_B$ at a temperature of 4.5 K---specifically, $n = (e^{\beta E_B} -1)^{-1} \approx 0.00024$.

	\newpage
	\section{S6: Emission time-traces vs dressing strength}
\begin{figure}[htp]
		\centering
		\includegraphics[width=4in]{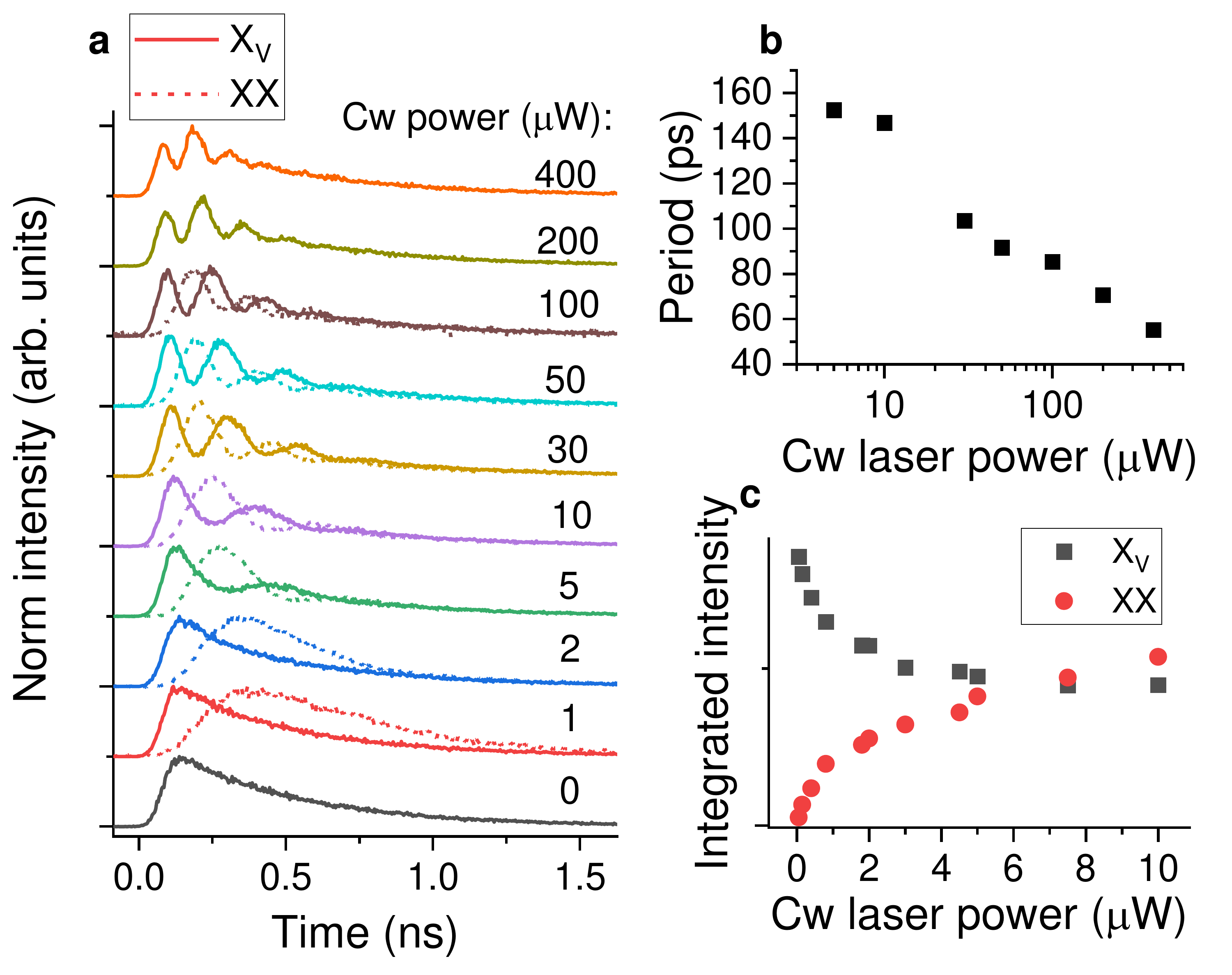}
		\caption{\label{fig:TRPLall}
			\textbf{Rabi oscillations in $X_V$ and $XX$ emission time-traces.} a,~Time-traces of the exciton and biexciton transitions as a function of cw dressing laser power resonant with $XX$-$X_V$ transition. b,~Period of Rabi oscillations vs cw dressing laser power in semi-log scale. c,~Integrated intensity of the exciton and biexciton emission vs cw dressing laser power. 
		}
	\end{figure}

\begin{figure}[htp]
		\centering
		\includegraphics[width=3.0in]{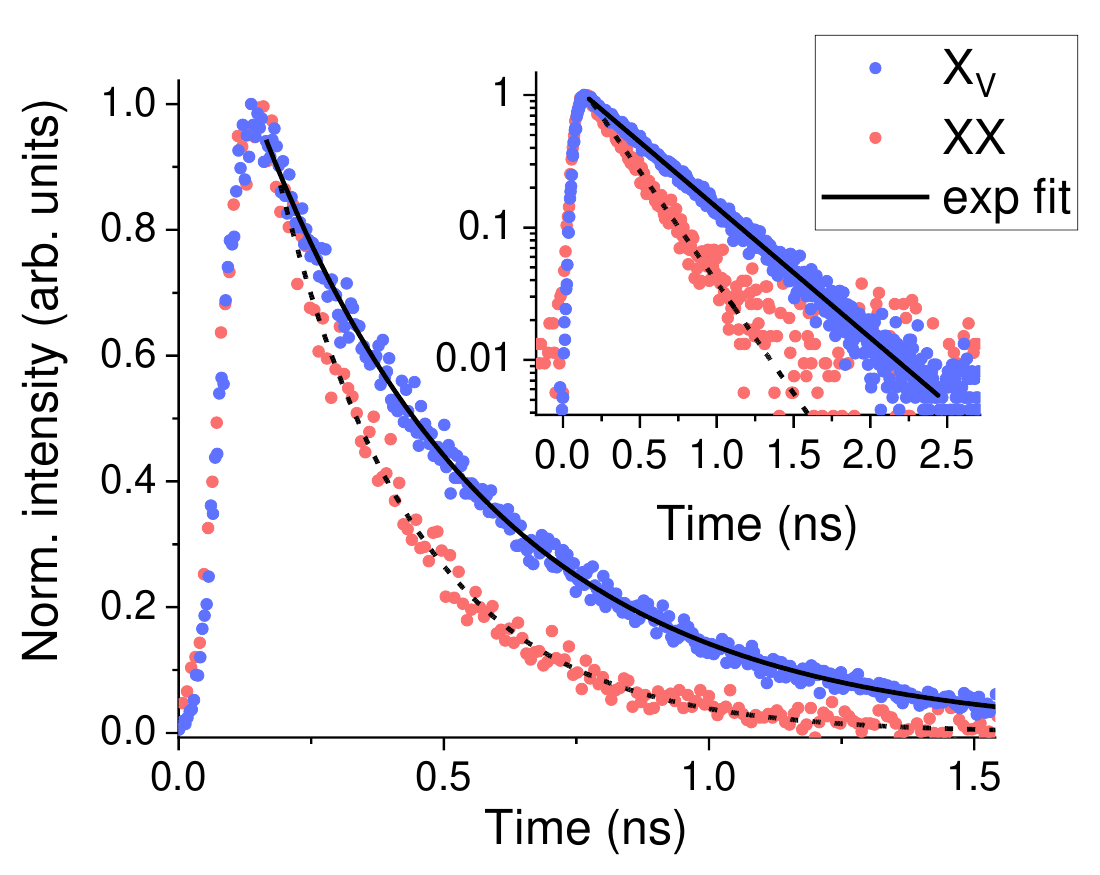}
		\caption{\label{fig:TRPLXX}
			\textbf{Bare biexciton and exciton emission time-traces.} Normalized emission time-trace of the bare exciton (blue points) and biexciton (red points) transitions under $\pi$-pulse excitation, exhibiting mono-exponential decays with 500~ps ($\hbar\gamma_X = 1.32 \ \mu$eV) and 260~ps ($\hbar\gamma_{XX} = 2.53 \ \mu$eV) time constants, respectively. The exciton decay is recorded under resonance fluorescence, while the biexciton is excited via resonant two-photon pumping. Inset: Semi-log plot of the corresponding graph.   	
		}
	\end{figure}

	In Fig.~\ref{fig:TRPLall}a, exciton and biexciton emission time-traces are shown for various dressing laser powers. For low excitation powers, the emission from the $\ket{XX}$-$\ket{X_V}$ transition can be clearly distinguished from the driving laser (dashed decay curves). At very intense $XX$ pumping, suppression of the dressing laser signal is not possible. The bare biexciton emission time-trace recorded under resonant two-photon excitation is shown in Fig.~\ref{fig:TRPLXX}.
	
	\newpage
	\section{S7: Correlation data analysis}
	For the $g^{(2)}[0]$ evaluation, the zero delay peak counts were integrated with a 4~ns time window and divided by the mean value of the neighbouring ten peaks' integrated counts. The uncertainty of $g^{(2)}[0]$ is based on the standard deviation of the non-central peaks' integrated counts. Correlation histograms were fitted with two-sided exponential decays convolved with a 350~ps width Gaussian instrumental response function (solid lines). In the dressed case, the correlation histogram exhibits a slight bunching feature on the 100~ns timescale which is accounted for in the fitting function. For the two-photon interference visibility evaluation, the experimental data have been fitted with two-side exponential decay functions, and the zero delay peak area calculated with respect to the neighbouring two peaks, with an additional offset to account for the residual coincidence counts due to cw excitation (see SI Sections 4 and 5). Next, the obtained raw values of 0.945$\pm$0.005 and 0.910$\pm$0.005 for bare and dressed case, respectively, have been corrected for the imperfections of the  Mach-Zehnder interferometer used~\cite{Dusanowski2019,Santori2002}, namely the beam-splitting ratio ($R/T$~=~1.15) and interferometer fringe contrast (1-$\varepsilon$~=~0.99), yielding the correction formula $\mathcal{V}_{\rm corr}=1.0303 \mathcal{V}_{\rm raw}$. To fit experimental data in Fig.4d, theoretically simulated decays based on Eq.(2) were additionally convolved with a 100~ps width Gaussian instrumental response function.
	
	\newpage
	\bibliography{bib-supplement}% Produces the bibliography via BibTeX.content...